\documentclass[11pt]{article}
\usepackage{authblk}
\usepackage{amsmath}
\usepackage{graphicx}
\newcommand{\bUp}{\mbox{\boldmath$\Upsilon$}}

\newcommand{\bSig}{\mbox{\boldmath$\Sigma$}}

\newcommand{\bB}{\mathbf{B}}
\newcommand{\bDc}{\mathbf{D}_{\mbox{\sf\tiny C}}}
\newcommand{\bDa}{\mathbf{D}_{\mbox{\sf\tiny A}}}

\newcommand{\bI}{\mathbf{I}}

\newcommand{\bT}{\mathbf{T}}
\newcommand{\bUc}{\mathbf{U}_{\mbox{\sf\tiny C}}}
\newcommand{\bUa}{\mathbf{U}_{\mbox{\sf\tiny A}}}
\newcommand{\tT}{\mbox{\sf\tiny (T)}}
\newcommand{\tC}{\mbox{\sf\tiny (C)}}
\newcommand{\tA}{\mbox{\sf\tiny (A)}}
\newcommand{\sT}{\mathsf{T}}

\newcommand{\qed}{{\sqcap \hspace*{-2.5mm}\sqcup}}
\title{\vspace*{-1.2in}
{\sf Monte Carlo Sampling Bias\\ in the Microwave Uncertainty Framework 
\footnote{\scriptsize Official contribution of the National Institute of Standards and Technology; not subject to copyright in the United
States.}}}
\author[1]{Michael Frey}
\author[2]{Benjamin F. Jamroz}
\author[1]{Amanda Koepke}
\author[2]{Jacob D. Rezac}
\author[2]{Dylan Williams}
\affil[1]{\small Statistical Engineering Division,
        National Institute of Standards and Technology,
       Boulder, Colorado 80305, USA}
\affil[2]{Radio Frequency Division,
        National Institute of Standards and Technology,
       Boulder, Colorado 80305, USA}
\date{\today}

\begin{document}
\maketitle

\begin{abstract}
Uncertainty propagation software can have unknown, inadvertent biases introduced by various means. This work is a case
study in bias identification and reduction in one such software package, the Microwave Uncertainty Framework (MUF).
The general purpose of the MUF is to provide automated multivariate statistical
uncertainty propagation and analysis on a Monte Carlo (MC) basis. {\sf Combine} is a key module in the MUF, responsible
for merging data, raw or transformed, to accurately reflect the variability in the data and in its central tendency. In this
work the performance of {\sf Combine}'s MC replicates is analytically compared against its stated design goals. An
alternative construction is proposed for {\sf Combine}'s MC replicates and its performance is compared, too, against
{\sf Combine}'s design goals. These comparisons are made within an archetypal two-stage scenario in which received
data are first transformed in conjunction with shared systematic error and then combined to produce summary information.
These comparisons reveal the limited conditions under which {\sf Combine}'s uncertainty results are unbiased and the
extent of these biases when these conditions are not met. For small MC sample sizes neither construction, current or
alternative, fully meets {\sf Combine}'s design goals, nor does either construction consistently outperform the other.
However, for large MC sample sizes the bias in the proposed alternative construction is asymptotically zero, and this
construction is recommended.
   
\vspace*{2mm}
\noindent
{\bf Keywords}: Monte Carlo sampling, sampling bias, statistical software,
statistical uncertainty propagation, systematic error, statistical experiment
\vspace*{2mm}
\end{abstract}

\section{Introduction}

Modern national, industrial, and academic laboratories engaged in high-precision metrology rely on statistical software for
multivariate and functional measurement uncertainty propagation and analysis. This software is typically highly complex
and flexible and often has a Monte Carlo basis. Even in software well-designed from a statistical perspective, biases can
be inadvertantly introduced due variously to flaws in statistical procedures, the algorithms that support them, or the
algorithms' coding. Statistical experiments are a natural and powerful way to test for such biases. We report the results
of a case study of a microwave measurement uncertainty software package, called the Microwave Uncertainty Framework
(MUF), in which a significant heretofore unknown bias in the software was detected, characterized, and corrected. This
case study shows that elementary statistical performance testing can successfully identify such biases.

\vspace*{2mm}
State-of-the-art microwave measurement relies on high-speed instrumentation including vector network analyzers (VNAs)
operating in the frequency domain, temporal sampling oscilloscopes, and an array of other instruments, often used
simultaneously in the same experiment. The refined measurements made possible by these arrangements allow
investigators to, for example, identify the multiple reflections created by small imperfections in microwave systems, capture
distortions due to the systems' frequency-limited electronics, and study the role of noise. Statistical analysis of the data from this mix
of instrumentation, including the conduct of uncertainty analyses, often involves shifts between the time and frequency domains. These
shifts require that microwave uncertainty analyses account, particularly, for statistical correlations among the measurement uncertainties.
To see this, consider that imperfections in microwave systems are often the source of unwanted reflections and attendant power
losses. These temporal effects Fourier-map into the frequency domain as ripples with a characteristic period related to the inverse
of the reflections' time spacing. The VNA is currently the most accurate instrument for measuring these multiple reflections,
and the errors made by this frequency sampling instrument typically manifest themselves as correlated time domain errors
in the magnitudes, shapes, and positions of the multiple reflections. Statistical uncertainties in VNA measurements cannot be
transformed correctly into the time domain without accounting for correlations created by the domain transformation \cite{hale}.

\vspace*{2mm}
Microwave measurement instrumentation with its often voluminous data production and the data-analytic need to track statistical
correlations have motivated three automated approaches for statistical uncertainty analysis, the METAS VNA Tools II software
package \cite{woll}, the Garelli-Ferrero (GF) software package \cite{gare}, and the MUF. The METAS and GF software packages
support microwave multi-port network investigations, and offer fast, efficient sensitivity analysis implementations \cite{avol}.

\vspace*{2mm}
The MUF is a software suite created, supported, and made publicly available by the Radio Frequency Division of the
U.S. National Institute of Standards and Technology. The MUF has capabilities similar to those of the METAS and GF software
packages, while supporting radio frequency engineering applications beyond network analysis. The MUF's general purpose is to
provide automated multivariate statistical type A- and type B-evaluated \cite{gum} uncertainty propagation and analysis on a Monte
Carlo (MC) basis \cite{supp} accessible through user-friendly interfaces. The MUF's MC capability preserves non-Gaussian features
of measured multivariate microwave signals, identifying systematic biases, for example, in signal calibration and processing steps.
The MUF is composed of functional modules selectable by the user as needed for analyses. These include {\sf Model} modules to
flexibly represent microwave system elements. {{\sf Model} modules are useful, for example, for building calibration models,
and they can be  cascaded to represent increasingly complex systems. Other processing modules, termed here {\sf Transform}
modules, are available to perform  oscilloscope and receiver calibrations, Fourier transforms, and other user-defined custom
analytical transformations. {\sf Combine} is a key module in the MUF, responsible for merging data, raw or transformed,
to accurately reflect the variability in the data and in its central tendency. {\sf Combine} is designed to be used at any point
in extended analyses where repeated measurements must be merged. This flexibility is a powerful feature of the MUF.

\vspace*{2mm}
This paper presents an analysis of the {\sf Combine} module in the MUF. Because of the MUF's distributed, multi-user, multi-purpose 
nature, {\sf Combine} can be executed at different stages of uncertainty propagation analysis. We study a common use of
{\sf Combine} described by the two-stage scenario diagrammed in Fig.\ 1. In the first stage of this scenario, multivariate data are joined
with shared systematic error in a bank of {\sf Transform} modules and then, in the second stage, the transformed data are combined
within the {\sf Combine} module. {\sf Transform}'s outputs take the form of nominal values of a selected mathematical transformation
with associated uncertainties. {\sf Transform}'s outputs in this form allow us in our two-stage scenario to study and assess
{\sf Combine}'s operation in which its inputs with their associated uncertainties are used to produce a summary mean output with an
associated uncertainty. {\sf Combine} represents the uncertainty in the summary mean in various fashions but provides the most detail
in the form of a sample of MC replicates. Our analysis of {\sf Combine} focuses specifically on the bias in the mean and covariance
of these MC replicates. This analysis reveals that {\sf Combine}'s construction of MC replicates is fundamentally biased, and we
propose an alternative construction that effectively eliminates this bias.

\begin{figure}[t!]
\vspace*{-8mm}
\begin{center}
\includegraphics[width=4.5in,height=3.5in]{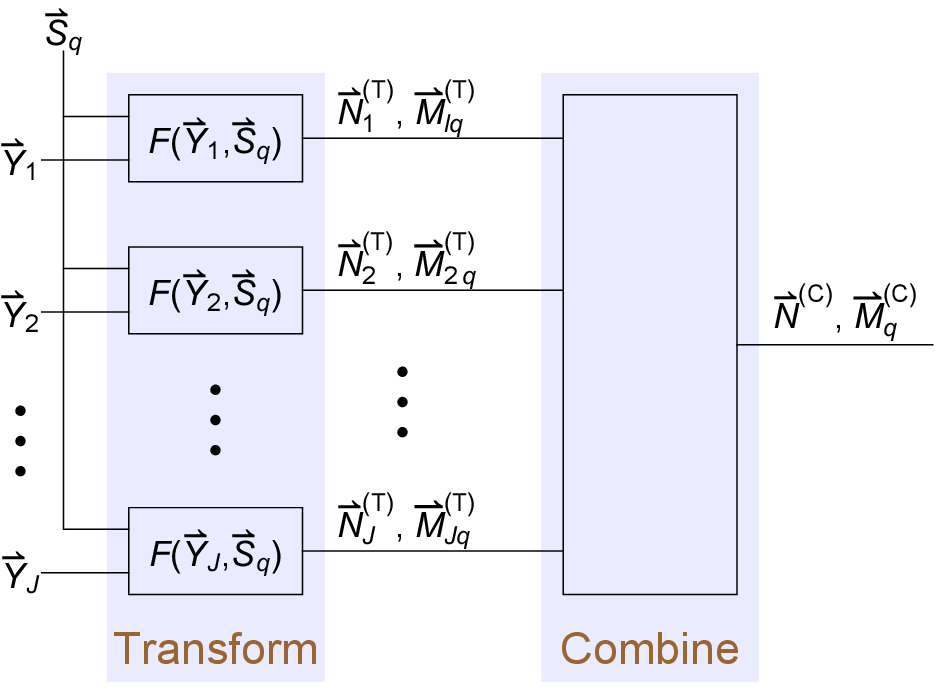}
\end{center}
\vspace*{1mm}
\begin{center}
\parbox{4.15in}{\small Figure 1. Two-stage MUF use scenario in which $J>1$ data vectors  $\vec{Y}_j$ are transformed in
conjunction with MC-generated shared systematic error $\vec{S}_q$ to produce nominal values $\vec{N}_j^{\tT}$ and
corresponding MC replicates $\vec{M}_{jq}^{\tT}$. These results are then merged in the {\sf Combine} module to
produce a combined nominal value $\vec{N}^{\tC}$ with MC replicates $\vec{M}_q^{\tC}$.}
\end{center}
\vspace*{-3mm}
\end{figure}

\vspace*{2mm}
The remainder of the paper is organized as follows. In Sect.\ 2 we analyze {\sf Combine}'s performance in the two-stage
scenario diagrammed in Fig.\ 1, showing that the sample mean of the MC replicates has zero bias and giving an analytical expression
for the bias in the covariance of its MC replicates. This covariance bias is studied for specific cases of additive, multiplicative,
exponential, and phase error. In Sect.\ 3 we propose an alternative construction for {\sf Combine}'s MC replicates and show
that the sample mean of {\sf Combine}'s MC replicates has zero bias. We put tight bounds on the corresponding covariance bias
and show that this bias is asymptotically zero; in this latter regard the proposed construction is better than the current method.
Estimation bias is the primary concern in MC sampling, but MC estimation variability is also an issue. In Sect.\ 4 we continue
our comparison of the current and alternative MC replicate constructions, comparing the variability in their sample means and
sample covariances. We conclude in Sect.\ 5 with summary remarks supporting adoption of the proposed alternative MC replicate
construction method in place of {\sf Combine}'s current method. For the results presented in the following sections, we assume
without note that the usual technical conditions pertain, that functions are measurable, that moments of sufficient order exist, etc.

\section{Bias in the {\sf Combine} module}

We suppose in the two-stage scenario in Fig.\ 1 that the $J>1$ data vectors $\vec{Y}_j$ (of length $K$) are identically distributed
and mutually independent and write $\vec{Y}_j\sim(\vec{\mu},\bSig)$ to identify the mean $\vec{\mu}$ and covariance
matrix $\bSig$ of $\vec{Y}_j$. We also suppose that the MC-generated, length-$K$ errors $\vec{S}_q$, $q=1,...,Q$ are
identically distributed, mutually independent, and independent of the sample of data vectors $\vec{Y}_j$. The mean and
covariance of $\vec{S}_q$ are $\vec{S}_q\sim(\vec{\nu},\bUp)$. The covariance $\bSig$ represents random uncertainty
in the measurement of $\vec{\mu}$ while the errors $\vec{S}_q$ are systematic post-measurement errors introduced
among the $\vec{Y}_j$ due, for example, to calibration adjustments.

\vspace*{2mm}
Each data vector $\vec{Y}_j$ in Fig.\ 1 is operated on individually by {\sf Transform}, producing for $\vec{Y}_j$ a nominal value
\begin{equation}
\vec{N}_j^{\tT}=F(\vec{Y}_j,\vec{\nu})
\label{eq:nom1}
\end{equation}
\noindent
for $j=1,..., J$ and a sample of vector Monte Carlo (MC) replicates
\begin{equation}
\vec{M}_{jq}^{\tT}=F(\vec{Y}_j,\vec{S}_q)
\label{eq:mc1}
\end{equation}
\noindent
for $q=1,..., Q$. The superscripts in (\ref{eq:nom1}) and (\ref{eq:mc1}) signify that these are {\sf Transform} outputs
in the first stage in our scenario. The MC replicates in (\ref{eq:mc1}) vary for a given $\vec{Y}_j$ only according to
random replicates from the distribution of $\vec{S}_q$. The same $Q$ random replicates $\vec{S}_q$ are used to
create each $\vec{Y}_j$'s sample of MC replicates. This models systematic errors that are shared among the $\vec{Y}_j$.
The {\sf Transform} module can similarly implement unshared systematic errors by using independent sets of $\vec{S}_q$
for each $\vec{Y}_j$, but the need for this capability rarely arises in application.

\vspace*{2mm}
The transformation $F$ in {\sf Transform} has the general form
\begin{equation}
F(\vec{y},\vec{s})=\left( \begin{matrix}
f(y_1,s_1) \\ \vdots \\  f(y_K,s_K)
\end{matrix}\right)
\label{eq:trans}
\end{equation}
\noindent
where $y_k$ and $s_k$ are the $k$th components of the vectors $\vec{y}$ and $\vec{s}$, respectively. The scalar-valued function
$f$ is a user-specified parameter in {\sf Transform}. Some choices of $f$ are $f(y,s)=y+s$, $f(y,s)=ys$, $f(y,s)=\sin(y+s)$, and
$f(y,s)=y^s$, representing additive, multiplicative, phase, and exponential error, respectively. {\sf Transform} also has many
optional parameters, among them two matrix parameters, $\bT_Y$ and $\bT_S$. When, for example, the user specifies a matrix
value for $\bT_Y$, $F$ is applied to $\bT_Y \vec{Y}$ instead of $\vec{Y}$. The matrix parameter $\bT_S$ operates similarly.
Mathematically, $\bT_Y$ and $\bT_S$ in the specification of $F$ are redundant; for example, we can without loss of generality
take $\bT_Y=\bI$ by substituting $\bT_Y \vec{\mu}$ and $\bT_Y\bSig\bT_Y^{\sT}$ for $\vec{\mu}$ and $\bSig$, respectively.
The optional use of $\bT_Y$, though, gives {\sf Transform} representational flexibility, allowing it, for example, to implement
the Fourier transform of $\vec{Y}_j$. The covariance $\bSig$ of $\vec{Y}_j$ is interpreted as the error covariance associated
with measurement of $\vec{\mu}$, so $\bSig$ and $\bT_Y$ reflect unrelated physical processes and have distinct modeling
roles. The matrices $\bUp$ and $\bT_S$ play similarly distinct modeling roles.

\vspace*{2mm}
{\sf Combine} in Fig.\ 1's two-stage scenario produces a nominal value $\vec{N}^{\tC}$ for the transformed data
and a sample of MC replicates $\vec{M}_q^{\tC}$ to describe the distribution and, particularly, the uncertainty of the 
transformed and combined data. These {\sf Combine} outputs are given by
\begin{equation}
\vec{N}^{\tC} = \frac{1}{J}\sum_{j=1}^J \vec{N}_j^{\tT} = \bar{F}(\vec{Y}_\bullet,\vec{\nu})
\label{eq:comN}
\end{equation}
\noindent
and
\begin{equation}
\vec{M}_q^{\tC}=\bar{M}_{\bullet q}^{\tT}+\frac{1}{\sqrt{J}} \bUc \sqrt{\bDc} \vec{Z}_q
\label{eq:comM}
\end{equation}
\noindent
where $\bar{M}_{\bullet q}^{\tT}=\bar{F}(\vec{Y}_\bullet,\vec{S}_q)$ and $\vec{Z}_q \sim \mbox{N}(\vec{0},\bI)$, 
$q=1,...,Q$. Further, the $\vec{Z}_q$ are independent of both the $\bar{M}_{\bullet q}^{\tT}$ and
$\bUc \sqrt{\bDc}$. The matrices $\bUc$ and
$\bDc$ are the unitary and diagonal members, respectively, of the eigendecomposition
$\hat{\bSig}_{\vec{N}_j^{\tT}} =\bUc \bDc \bUc^{\sT}$ of the sample covariance matrix
\begin{equation}
\hat{\bSig}_{\vec{N}_j^{\tT}}
= \frac{1}{J-1} \sum_{j=1}^J (\vec{N}_j^{\tT}-\bar{N}_\bullet^{\tT})
(\vec{N}_j^{\tT}-\bar{N}_\bullet^{\tT})^\sT 
\end{equation}
\noindent
associated with the nominal vectors $\vec{N}_j^{\tT}$ at {\sf Combine}'s input.

\vspace*{2mm}
The vectors $\vec{Z}_q$ in (\ref{eq:comM}) model standard normal variation along the principal axes of
$\hat{\bSig}_{\vec{N}_j^{\tT}}$. These standard normal variates are scaled by the standard deviations in $\sqrt{\bDc}$
and then rotated by $\bUc$ onto the coordinate axes of $\hat{\bSig}_{\vec{N}_j^{\tT}}$ to reflect the variability of the
transformed data at {\sf Combine}'s input. The components of the $\vec{Z}_q$ are chosen to be normally distributed
based on the assumption that the number $J$ of {\sf Combine} inputs and their independence are together great enough
to support a Central Limit Theorem approximation. We present the $\vec{Z}_q$ as normally distributed because this is how
they are generated in {\sf Combine}. Only in subsection 4.2, however, is this distributional assumption necessary to our results.

\vspace*{2mm}
The nominal value $\vec{N}^{\tC}$ in (\ref{eq:comN}) produced by {\sf Combine} is a natural, intuitive summary of the central
tendency of the transformed data provided that the transformed data are unimodal with little skew. The purpose of the
MC replicates $\vec{M}_q^{\tC}$ is to indicate central tendency under more general conditions as well as to
summarize the spread and distributional shape of the estimated central tendency. Formally, {\sf Combine} is designed
to produce a sample of MC replicates whose mean $\bar{M}_\bullet^{\tC}$ is an unbiased estimator of the vector
$E[\bar{F}(\vec{Y}_\bullet,\vec{S}_q)]$ and whose covariance
\begin{equation}
\hat{\bSig}_{\vec{M}_q^{\tC}} = \frac{1}{Q-1} \sum_{q=1}^Q (\vec{M}_q^{\tC}-\bar{M}_\bullet^{\tC})
(\vec{M}_q^{\tC}-\bar{M}_\bullet^{\tC})^\sT
\end{equation}
\noindent
is an unbiased estimator of the covariance $V[\bar{F}(\vec{Y}_\bullet,\vec{S}_q)]$ of the vector
$\bar{F}(\vec{Y}_\bullet,\vec{S}_q)$. In other words, the MC replicates in (\ref{eq:comM}) should satisfy
\begin{equation}
E[\bar{M}_\bullet^{\tC}] = E[\bar{F}(\vec{Y}_\bullet,\vec{S}_q)]
\label{eq:cond1}
\end{equation}
and
\noindent
\begin{equation}
E[\hat{\bSig}_{\vec{M}_q^{\tC}}] = V[\bar{F}(\vec{Y}_\bullet,\vec{S}_q)] .
\label{eq:cond2}
\end{equation}
\noindent
We note for later use that under the conditions of our two-stage scenario the estimands
in (\ref{eq:cond1}) and (\ref{eq:cond2}) can be expressed as
\begin{equation}
E[\bar{F}(\vec{Y}_\bullet,\vec{S}_q)] = E[F(\vec{Y}_j,\vec{S}_q)]
\label{eq:simp1}
\end{equation}
and
\noindent
\begin{eqnarray}
V[\bar{F}(\vec{Y}_\bullet,\vec{S}_q)] \hspace*{-6mm}
&& = \frac{1}{J} V[F(\vec{Y}_j,\vec{S}_q)]   \nonumber \\[-1mm]
&& \label{eq:simp2} \\[-3mm]
&& \hspace*{-6mm} +\,\frac{J-1}{J} Cov[F(\vec{Y}_j,\vec{S}_q),F(\vec{Y}_{j^{\prime}},\vec{S}_q)] \nonumber
\end{eqnarray}
\noindent
with $j\neq j^\prime$. Our analysis, summarized in Proposition 1 below, of the MC construction in (\ref{eq:comM})
shows that {\sf Combine} meets design goals (\ref{eq:cond1}) and (\ref{eq:cond2}) only under certain conditions,
and that without these conditions {\sf Combine} exhibits bias.

%--------------------------------------------------------------------------------------------------------------- Proposition 1
\vspace*{2mm}
{\it Proposition 1}: Suppose that, in the two-stage scenario in Fig.\ 1 , we have $J>1$ independent, identically distributed data
vectors $\vec{Y}_j\sim(\vec{\mu},\bSig)$. Also suppose we have $Q$ independent, identically distributed errors
$\vec{S}_q\sim(\vec{\nu},\bUp)$. Assume the sets of $\vec{Y}_j$ and $\vec{S}_q$ are independent. Suppose further that the
{\sf Transform} outputs $\vec{N}_j^{\tT}=F(\vec{Y}_j,\vec{\nu})$ and $\vec{M}_{jq}^{\tT}$ are given by (\ref{eq:nom1})
and (\ref{eq:mc1}) with $F$ as in (\ref{eq:trans}), and the {\sf Combine} outputs $\vec{N}^{\tC}$ and
$\vec{M}_q^{\tC}$ are given by (\ref{eq:comN}) and (\ref{eq:comM}). Then
\begin{equation}
E[\bar{M}_\bullet^{\tC}] = E[\bar{F}(\vec{Y}_\bullet,\vec{S}_q)]
\label{eq:th1}
\end{equation}
and
\noindent
\begin{equation}
E[\hat{\bSig}_{\vec{M}_q^{\tC}}] = V[\bar{F}(\vec{Y}_\bullet,\vec{S}_q)]+\frac{1}{J}\Psi
\label{eq:th2}
\end{equation}
\noindent
where $\Psi$ is the difference of two $K\!\times\! K$ covariances
\begin{equation}
\Psi = V[F(\vec{Y}_j,\vec{\nu})] - V[E[F(\vec{Y}_j,\vec{S}_q)|\vec{Y}_j]] .
\label{eq:Psi}
\end{equation}

\vspace*{2mm} Proposition 1 establishes that the design goal in (\ref{eq:cond1}) is generally met by the MC
replicates in (\ref{eq:comM}), but the design goal in (\ref{eq:cond2}) is not. The covariance in the sample
of MC replicates is biased by an amount $\Psi/J$. We will see in the next section that this bias
can be positive or negative. We first prove the two parts (\ref{eq:th1}) and (\ref{eq:th2}) of the proposition.

%--------------------------------------------------------------------------------------------------------------- Proof of (12)
\vspace*{2mm}
{\it Proof of (\ref{eq:th1})}: We first note that $E[\bar{M}_\bullet^{\tC}] = E[\vec{M}_q^{\tC}]$ since the $\vec{M}_q^{\tC}$
are identically distributed. Then, conditioning on the factor $\bUc \sqrt{\bDc}$ in
(\ref{eq:comM}) and using that $\vec{Z}_q$ and $\bUc \sqrt{\bDc}$ are independent, we have
\begin{eqnarray}
E[\bar{M}_\bullet^{\tC}]  \hspace*{-6mm}
&& = E[E[\bar{M}_{\bullet q}^{\tT}+\frac{1}{\sqrt{J}} \bUc \sqrt{\bDc} \vec{Z}_q|\bUc \sqrt{\bDc}]]
\nonumber \\
&& =  E[E[\bar{M}_{\bullet q}^{\tT}|\bUc \sqrt{\bDc}]] +\frac{1}{\sqrt{J}}
E[\bUc \sqrt{\bDc}]E[ \vec{Z}_q] . \nonumber
\end{eqnarray}
\noindent
Since $E[ \vec{Z}_q]=\vec{0}$, this yields $E[\bar{M}_\bullet^{\tC}] = E[\bar{M}_{\bullet q}^{\tT}]
= E[\bar{F}(\vec{Y}_\bullet,\vec{S}_q)]$, which proves (\ref{eq:th1}). \hfill $\qed$

\vspace*{2mm}
To prove (\ref{eq:th2}) in the proposition, we need four lemmas, which we state here. Their proofs are given in the appendix.
Lemma 1 concerns the sample covariance of cross-correlated vectors. Lemmas 2 and 3 are elementary conditioning argument-based
results for auto- and cross-covariances. Lemma 4 is used here and in the proofs of subsequent propositions.

\vspace*{2mm}
{\it Lemma 1}: Let $\vec{X}_j\sim (\vec{\mu},\bSig)$ for $j=1,...,J$, $J>1$ with sample covariance
\begin{equation}
\hat{\bSig}=\frac{1}{J-1} \sum_{j=1}^J (\vec{X}_j-\bar{X}) (\vec{X}_j-\bar{X})^\sT .
\nonumber
\end{equation}
\noindent
Let $Cov[\vec{X}_j,\vec{X}_k]=E[(\vec{X}_j-E[\vec{X}_j])(\vec{X}_k-E[\vec{X}_k])^\sT]$ be the cross-covariance of $\vec{X}_j$
and $\vec{X}_k$, and suppose the $\vec{X}_j$ are cross-correlated with $Cov[\vec{X}_j,\vec{X}_k]=\bSig^\prime$
for all $j\neq k$. Then $E[\hat{\bSig}]=\bSig-\bSig^\prime$.

\vspace*{2mm}
{\it Lemma 2}: Let $\vec{Z}\sim(\vec{0},\bI)$ be independent of the vector-matrix pair $(\vec{A},\bB)$. Then
$V[\vec{A}+\bB\vec{Z}]=V[\vec{A}]+E[\bB\bB^\sT]$.

\vspace*{2mm}
{\it Lemma 3}: Let $\vec{Z}_1,\vec{Z}_2\sim(\vec{0},\bI)$, and suppose $\vec{Z}_1$, $\vec{Z}_2$, and  $(\vec{A}_1,\vec{A}_2,\bB)$
are mutually independent. Then $Cov[\vec{A}_1+\bB\vec{Z}_1,\vec{A}_2+\bB\vec{Z}_2]=Cov[\vec{A}_1,\vec{A}_2]$.

\vspace*{2mm}
{\it Lemma 4}: Let $\vec{S},\vec{S}^\prime$ be independent, identically distributed random vectors independent of the
random vector $\vec{Y}$. Let $F(\vec{Y},\vec{S})$ be a vector function of $\vec{Y}$ and $\vec{S}$. Then
$Cov[F(\vec{Y},\vec{S}),F(\vec{Y},\vec{S}^\prime)] = V[E[F(\vec{Y},\vec{S})|\vec{Y}]]$.

%--------------------------------------------------------------------------------------------------------------- Proof of (13)
\vspace*{2mm}
{\it Proof of (\ref{eq:th2})}: The MC replicate vectors $\vec{M}_q^{\tC}$ created by {\sf Combine} are correlated with
common cross-covariance $Cov[\vec{M}_q^{\tC},\vec{M}_{q^\prime}^{\tC}]$. Therefore, according to Lemma 1,
\begin{equation}
E[\hat{\bSig}_{\vec{M}_q^{\tC}}] = V[\vec{M}_q^{\tC}]-Cov[\vec{M}_q^{\tC},\vec{M}_{q^\prime}^{\tC}] 
\label{eq:stp1}
\end{equation}
\noindent
with $q\neq q^\prime$. Using definition (\ref{eq:comM}) for $\vec{M}_q^{\tC}$, Lemma 2, definition (\ref{eq:mc1}) for
$\vec{M}_{jq}^{\tT}$, and the eigendecomposition $\hat{\bSig}_{\vec{N}_j^{\tT}} =\bUc \bDc \bUc^{\sT}$, we have
\begin{eqnarray}
V[\vec{M}_q^{\tC}] \hspace*{-6mm} && = V\left[ \bar{M}_{\bullet q}^{\tT}
+ \frac{1}{\sqrt{J}} \bUc\sqrt{\bDc}\vec{Z}_q\right]
\nonumber \\
&& =V[ \bar{M}_{\bullet q}^{\tT}] +  \frac{1}{J} E[\bUc \bDc \bUc^\sT ] \nonumber \\
&& =V[\bar{F}(\vec{Y}_\bullet,\vec{S}_q)] +  \frac{1}{J} E[\hat\bSig_{\vec{N}_j}^{\tT} ] . \label{eq:int1}
\end{eqnarray}
\noindent
The {\sf Transform} nominal values $\vec{N}_j^{\tT}$ in (\ref{eq:nom1}) are independent and identically distributed so
\begin{equation}
E[\hat\bSig_{\vec{N}_j}^{\tT} ]= V[\vec{N}_j^{\tT}]= V[F(\vec{Y}_j,\vec{\nu})]
\label{eq:resES1}
\end{equation}
\noindent
and (\ref{eq:int1}) becomes
\begin{equation}
V[\vec{M}_q^{\tC}]=V[\bar{F}(\vec{Y}_\bullet,\vec{S}_q)] +  \frac{1}{J}  V[F(\vec{Y}_j,\vec{\nu})] .
\label{eq:ins1}
\end{equation}

\vspace*{2mm}
Now consider the cross-covariance $Cov[\vec{M}_q^{\tC},\vec{M}_{q^\prime}^{\tC}]$ in (\ref{eq:stp1}). Using
definition (\ref{eq:comM}) for $\vec{M}_q^{\tC}$, Lemma 3, and definition (\ref{eq:mc1}) for
$\vec{M}_{jq}^{\tT}$, we have
\begin{eqnarray}
Cov[\vec{M}_q^{\tC},\vec{M}_{q^\prime}^{\tC}]  \hspace*{-6mm}
&& = Cov\! \left[ \bar{M}_{\bullet q}^{\tT} + \frac{1}{\sqrt{J}} \bUc\sqrt{\bDc}\vec{Z}_q,
\bar{M}_{\bullet q^\prime}^{\tT} + \frac{1}{\sqrt{J}} \bUc\sqrt{\bDc}\vec{Z}_{q^\prime} \right] \nonumber \\
&& = Cov[ \bar{M}_{\bullet q}^{\tT} , \bar{M}_{\bullet q^\prime}^{\tT} ] \nonumber \\
&& = \frac{1}{J^2} Cov\! \left[ \sum_{j=1}^J F(\vec{Y}_j,\vec{S}_q), \sum_{j^\prime=1}^J F(\vec{Y}_{j^\prime},\vec{S}_{q^\prime})
\right] \nonumber \\
&& = \frac{1}{J} Cov[F(\vec{Y}_j,\vec{S}_q), F(\vec{Y}_j,\vec{S}_{q^\prime})], \label{eq:int2}
\end{eqnarray}
\noindent
the last equality holding because the data vectors $\vec{Y}_j$ are independent.
Applying Lemma 4 to the covariance in (\ref{eq:int2}) and substituting the result along with (\ref{eq:ins1})
back into (\ref{eq:stp1}) proves (\ref{eq:th2}). \hfill $\qed$

\subsection{Example error models}

\vspace*{2mm}
Proposition 1's point is that the MC replicates produced by {\sf Combine} in our two-stage scenario have
a covariance bias $\Psi/J$. In the remainder of this section we evaluate $\Psi$ for various error models,
showing that $\Psi$ can be positive, negative, or zero. Where $\Psi$ is non-zero, we show in the
univariate case $K=1$ that the relative bias
\begin{equation}
\mbox{relbias}  [ \hat{\bSig}_{\vec{M}_q^{\tC}}]
= \frac{E[\hat{\bSig}_{\vec{M}_q^{\tC}}]-V[\bar{F}(\vec{Y}_\bullet,\vec{S}_q)]}{V[\bar{F}(\vec{Y}_\bullet,\vec{S}_q)]}
\label{eq:relbias}
\end{equation}
\noindent
approaches $\pm$20\% in one example and even 200\% in another.

\vspace*{2mm}
{\it Additive error}: The function $f$ in (\ref{eq:trans}) is $f(a,b)=a+b$ for additive error. In this case
$F(\vec{y},\vec{s}) = \vec{y}+\vec{s}$ and
\begin{eqnarray}
E[F(\vec{Y}_j,\vec{S}_q)|\vec{Y}_j]  \hspace*{-6mm}
&&= E[\vec{Y}_j+\vec{S}_q|\vec{Y}_j] \nonumber \\
&&= \vec{Y}_j +  E[\vec{S}_q] \nonumber \\
&&= F(\vec{Y}_j,\vec{\nu}) \nonumber
\end{eqnarray}
\noindent
so $\Psi$ in (\ref{eq:Psi}) is identically zero. Thus for additive shared systematic error {\sf Combine}'s MC replicates
have both zero mean bias and zero covariance bias.

\vspace*{2mm}
{\it Multiplicative error}: The function $f$ in (\ref{eq:trans}) is $f(y,s)=ys$ for multiplicative error
and the $k$th component of $F(\vec{Y}_j,\vec{S}_q)$ is $f({Y}_{jk},{S}_{qk})=Y_{jk} S_{qk}$ where $Y_{jk}$
and $S_{qk}$ are the $k$th components of $\vec{Y}_j$ and $\vec{S}_q$. We have
\begin{eqnarray}
E[f(Y_{jk},S_{qk})|\vec{Y}_j]  \hspace*{-6mm}
&&= E[Y_{jk}S_{qk}|\vec{Y}_j] \nonumber \\
&&=Y_{jk} E[S_{qk}|\vec{Y}_j] \nonumber \\
&&=Y_{jk} \,\nu_k . \nonumber
\end{eqnarray}
\noindent
Therefore $E[F(\vec{Y}_j,\vec{S}_q)|\vec{Y}_j]=F(\vec{Y}_j,\vec{\nu})$ and $\Psi=0$. This shows that
for multiplicative shared systematic error {\sf Combine}'s MC replicates have both zero mean bias and zero covariance bias.

\vspace*{2mm}
{\it Phase error}: The function $f$ in (\ref{eq:trans}) is $f(y,s)=\sin(y+s)$ for phase error. In this case the covariance in
{\sf Combine}'s MC replicates can be biased. We focus on the univariate case $K=1$ in which we have scalars, $Y_j$ and
$S_q$, and the phase error is uniformly distributed, $S_q \sim \mbox{Unif}(-\delta,\delta)$, $\delta>0$, with mean
$\nu=0$ and range $2\delta$. We note first that
\begin{equation}
E[\sin(Y_j+S_q)|Y_j=y]=\frac{\sin y\sin\delta}{\delta} .
\nonumber
\end{equation}
\noindent
Therefore
\begin{equation}
V[E[\sin(Y_j+S_q)|Y_j]]=V\left[\frac{\sin Y_j\sin\delta}{\delta}\right] = \frac{\sin^2\delta}{\delta^2} V[\sin Y_j]
\nonumber
\end{equation}
\noindent
and, using $\nu= E[S_q] =0$,
\begin{eqnarray}
\Psi \hspace*{-6mm} && =V[\sin(Y_j+0)]-\frac{\sin^2\delta}{\delta^2} V[\sin Y_j] \nonumber \\
&& =\left( 1-\frac{\sin^2\delta}{\delta^2} \right) V[\sin Y_j] \nonumber \\
&& >0.      \nonumber
\end{eqnarray}

\vspace*{2mm}
To assess the relative size of the bias associated with $\Psi>0$ above, we consider the extremal case where $\delta=\pi$
and where $Y_j$ is $\pm\pi/2$ with equal probabilities. Then $\Psi = 1$, $V[\sin(Y_j+S_q)] = 1/2$, and
$Cov[\sin(Y_j+S_q),\sin(Y_{j^{\prime}}+S_q)] = 0$ for $j\neq j^\prime$. Using (\ref{eq:simp2}), we find that the relative bias
(\ref{eq:relbias}) associated with the MC sample variance is
\begin{displaymath}
\mbox{relbias}[ \hat{\bSig}_{\vec{M}_q^{\tC}} ]
= \frac{\frac{1}{J}\Psi}{V[\bar{F}(\vec{Y}_\bullet,\vec{S}_q)]}
= \frac{\frac{1}{J}1}{\frac{1}{J}\frac{1}{2}+\frac{J-1}{J}0} =2 .
\end{displaymath}  
\noindent
Here the relative bias is 200\% for any sample size $J$. This albeit exteme example demonstrates
that very large relative biases are possible with {\sf Combine}'s current method of MC replicate construction. 

\vspace*{2mm}
{\it Exponential error}: The function $f$ in (\ref{eq:trans}) is $f(y,s)=y^s$ for exponential error. In this case the covariance
in {\sf Combine}'s MC replicates can be positively or negatively biased. We focus on the case $K=1$ of uniformly distributed scalars,
$Y_j$ and $S_q$. For this case we find that $\Psi$ is broadly, but not always, negative.

\vspace*{2mm}
Let $Y_j\sim\mbox{Unif}[a,b]$ and $S_q\sim\mbox{Unif}[1-\alpha,1+\alpha]$, $0\leq \alpha\leq 1$. We have $\nu=E[S_q]=1$ so
$\Psi$ in (\ref{eq:Psi}) is $V[Y_j] - V[k(Y_j)]$ where
\begin{equation}
k(y) = E[y^{S_q}] = \frac{y \sinh(\alpha \ln y)}{\alpha \ln y} .
\nonumber
\end{equation}
\noindent
Using $V[Y_j] =\frac{1}{12}(b-a)^2$ and evaluating $V[k(Y_j)]$ numerically, we find that $\Psi$ is slightly positive for
small $b$, as shown in Fig.\ 2. Otherwise, in the region $0\leq a \leq b \leq 8$, $\Psi$ is
negative, increasingly so for larger ranges $2\alpha$ and $b-a$.

In the cases presented in Fig.\ 2 for exponential error, the covariance
\begin{displaymath}
Cov[F(\vec{Y}_j,\vec{S}_q),F(\vec{Y}_{j^{\prime}},\vec{S}_q)] = Cov[Y_j^{S_q},Y_{j^{\prime}}^{S_q}]
\end{displaymath}
\noindent
is positive. According to (\ref{eq:simp2}), then, the relative bias (\ref{eq:relbias}) associated with $\Psi$ is strongest at the smallest sample size $J=2$, in which case
\begin{displaymath}
\mbox{relbias}[ \hat{\bSig}_{\vec{M}_q^{\tC}}] = \frac{\Psi}{V[Y_j^{S_q}]+Cov[Y_j^{S_q},Y_{j^\prime}^{S_q}]} .
\end{displaymath} 
\noindent
Numerical evaluation of this expression yields the results presented in Fig.\ 3. At its strongest the relative bias
approaches $\pm$20\% for $\alpha = 0.95$.

\begin{figure}
\vspace*{-20mm}
\begin{center}
\includegraphics[width=2.8in,height=6.8in]{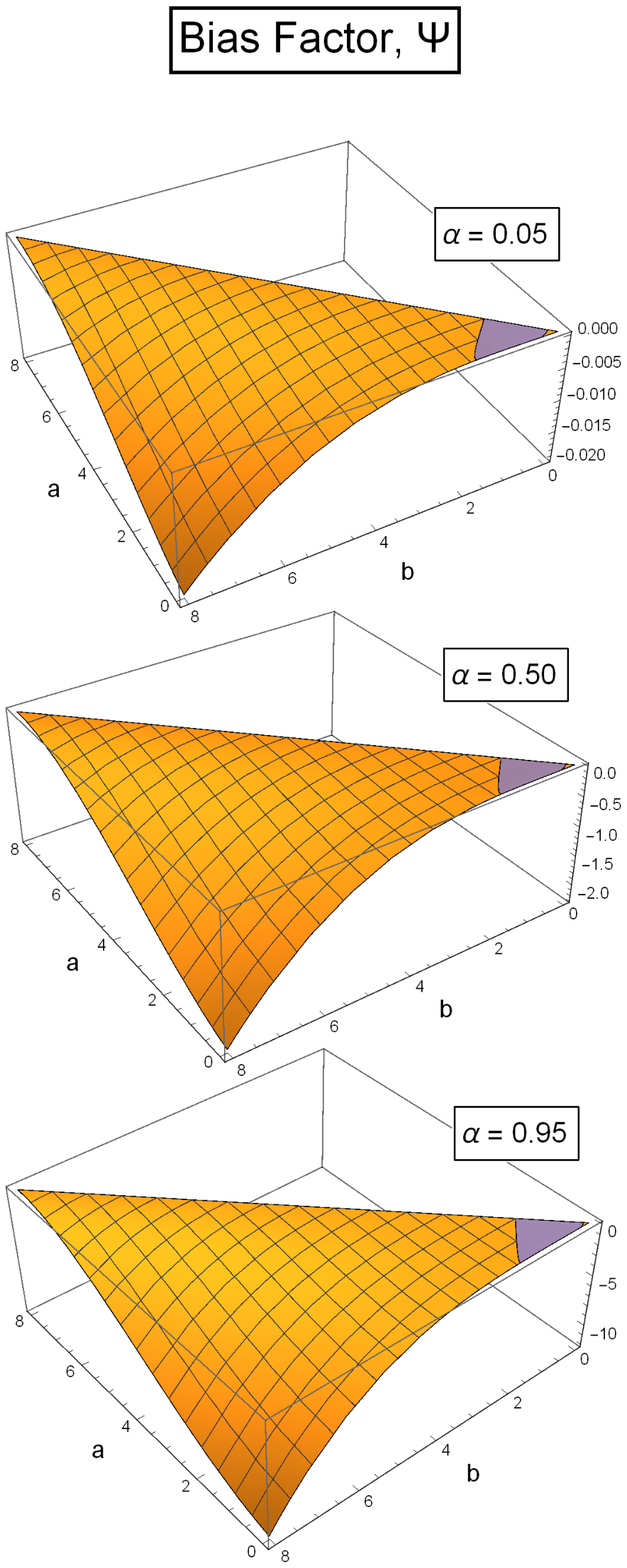}
\end{center}
\vspace*{-10mm}
\begin{center}
\parbox{4.1in}{\small Figure 2. Bias factor $\Psi$ with exponential error. The data $Y_j$ and the exponential error $S_q$
are both uniformly distributed, $Y_j\sim\mbox{Unif}[a,b]$ and $S_q\sim\mbox{Unif}[1-\alpha,1+\alpha]$. The bias factor
$\Psi$ is negative everywhere except in the small blue patches where $\Psi >0$.}
\end{center}
\end{figure}

\begin{figure}
\vspace*{-20mm}
\begin{center}
\includegraphics[width=2.8in,height=6.8in]{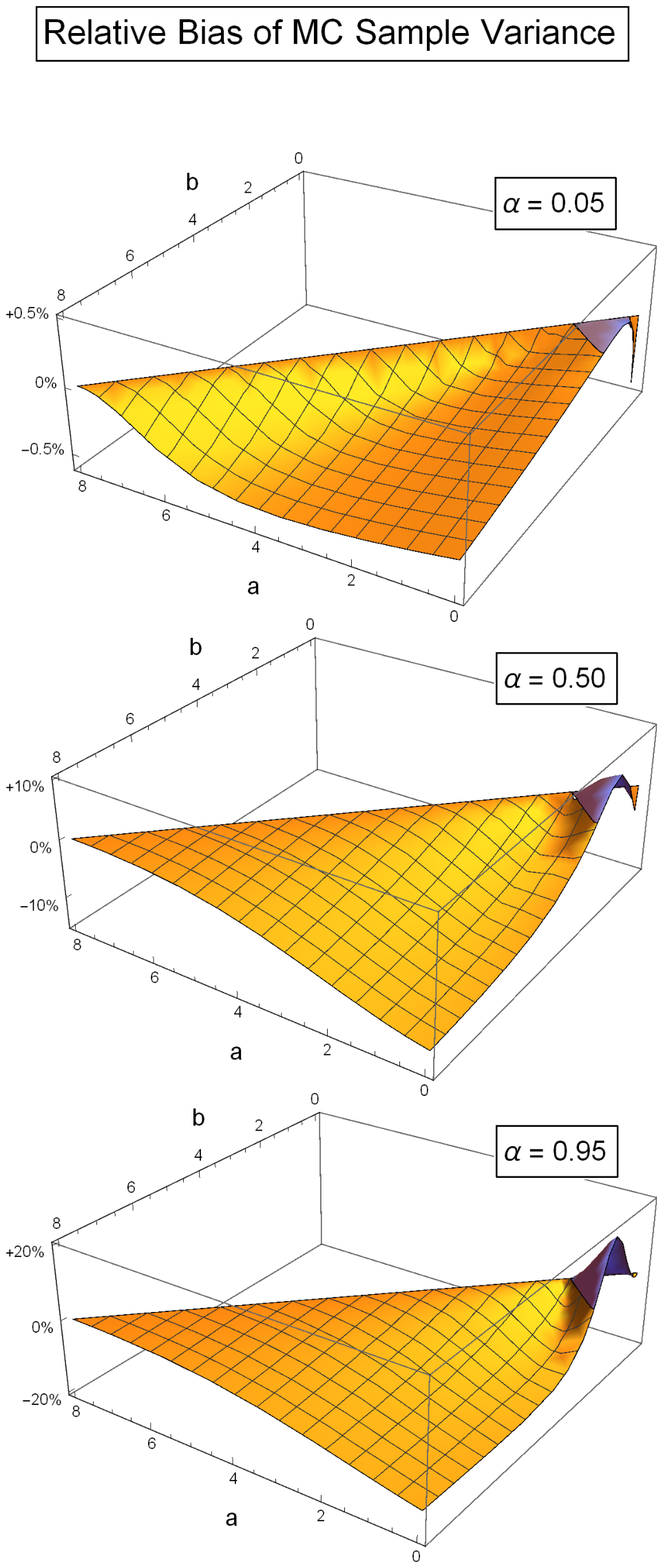}
\end{center}
\vspace*{-10mm}
\begin{center}
\parbox{4.1in}{\small Figure 3. Relative bias of {\sf Combine}'s MC sample variance with exponential error.
The data $Y_j$ and the exponential error $S_q$ are both uniformly distributed, $Y_j\sim\mbox{Unif}[a,b]$
and $S_q\sim\mbox{Unif}[1-\alpha,1+\alpha]$. The relative bias approaches $\pm$20\% for $\alpha = 0.95$.}
\end{center}
\end{figure}

\section{An alternative MC construction}

The previous section shows that {\sf Combine}'s MC replicates $\vec{M}_q^{\tC}$ in (\ref{eq:comM}) generated for the two-stage scenario in Fig.\ 1 fail to fully meet {\sf Combine}'s design goals (\ref{eq:cond1}) and (\ref{eq:cond2}). We propose in this section
an alternative construction $\vec{M}_q^{\tA}$ for {\sf Combine}'s MC replicates. Like the $\vec{M}_q^{\tC}$ replicates in
(\ref{eq:comM}), the proposed $\vec{M}_q^{\tA}$ replicates meet goal (\ref{eq:cond1}). Unlike the $\vec{M}_q^{\tC}$
replicates, the $\vec{M}_q^{\tA}$ replicates essentially meet goal (\ref{eq:cond2}), doing so arbitrarily closely for sufficiently
large MC replicate sample size $Q$.

\vspace*{2mm}
Let
\begin{equation}
\vec{M}_q^{\tA}=\bar{M}_{\bullet q}^{\tT}+\frac{1}{\sqrt{J}} \bUa \sqrt{\bDa} \vec{Z}_q
\label{eq:comM2}
\end{equation}
\noindent
where $\bar{M}_{\bullet q}^{\tT}=\bar{F}(\vec{Y}_\bullet,\vec{S}_q)$ and
$\vec{Z}_q \sim \mbox{N}(\vec{0},\bI)$, $q=1,...,Q$. Further, the $\vec{Z}_q$ are independent of both the
$\bar{M}_{\bullet q}^{\tT}$ and $\bUa \sqrt{\bDa}$. In this alternative construction
the matrices $\bUa$ and $\bDa$ are now the unitary and diagonal members, respectively, of the eigendecomposition
$\hat{\bSig}_{\bar{M}_{j\bullet}^{\tT}} =\bUa \bDa \bUa^{\sT}$ of the sample covariance
\begin{equation}
\hat{\bSig}_{\bar{M}_{j\bullet}^{\tT}}
= \frac{1}{J-1} \sum_{j=1}^J (\bar{M}_{j\bullet}^{\tT}-\bar{M}_{\bullet\bullet}^{\tT})
(\bar{M}_{j\bullet}^{\tT}-\bar{M}_{\bullet\bullet}^{\tT})^\sT 
\end{equation}
\noindent
associated with the means $\bar{M}_{j\bullet}^{\tT}$ of the MC samples at {\sf Combine}'s input. Proposition 2
below shows that basing the sample variability of the stage-two {\sf Combine} MC replicates on the stage-one MC
means $\bar{M}_{j\bullet}^{\tT}$ instead of on the stage-one nominal values $\vec{N}_j^{\tT}$ essentially
removes the bias $\Psi/J$ identified in Proposition 1. This reduced bias is explained in some part by the greater
information retained by using the MC means instead of the nominal values: the $M_{jq}^{\tT} = F(\vec{Y}_j,\vec{S}_q)$
reflect nonlinearities in $F$ across the full distribution of $\vec{S}$, while the nominal values
$\vec{N}_j^{\tT}=F(\vec{Y}_j,\vec{\nu})$ are only exposed to $F$ at the mean $\vec{\nu}$ of the $\vec{S}_q$ distribution. 

%--------------------------------------------------------------------------------------------------------------- Proposition 2

\vspace*{2mm}
{\it Proposition 2}: Let the set-up be the same as in Proposition 1 except that the {\sf Combine}-stage
MC replicates in Fig.\ 1 are given by $\vec{M}_q^{\tA}$ in (\ref{eq:comM2}). Then
\begin{equation}
E[\bar{M}_\bullet^{\tA}] = E[\bar{F}(\vec{Y}_\bullet,\vec{S}_q)]
\label{eq:th21}
\end{equation}
and
\noindent
\begin{equation}
E[\hat{\bSig}_{\vec{M}_q^{\tA}}] = V[\bar{F}(\vec{Y}_\bullet,\vec{S}_q)]+\frac{1}{JQ}\Phi 
\label{eq:th22}
\end{equation}
\noindent
where $\Phi$ is the difference of two $K\!\times\! K$ covariances
\begin{equation}
\Phi = E[V[F(\vec{Y}_j,\vec{S}_q)|\vec{S}_q]] - V[E[F(\vec{Y}_j,\vec{S}_q)|\vec{Y}_j]] .
\label{eq:Phi}
\end{equation}

%--------------------------------------------------------------------------------------------------------------- Proof
\vspace*{2mm}
{\it Proof}: The proof of (\ref{eq:th21}) is the same as that of (\ref{eq:th1}) because $\vec{Z}_q$ and
$\bUa \sqrt{\bDa}$ in (\ref{eq:comM2}) are again independent. To prove (\ref{eq:th22}), we first note that
the arguments based on Lemmas 1, 2, and 3 early in the proof of (\ref{eq:th2}) apply also here, giving
\begin{equation}
E[\hat{\bSig}_{\vec{M}_q^{\tA}}] = V[\vec{M}_q^{\tA}]-Cov[\vec{M}_q^{\tA},\vec{M}_{q^\prime}^{\tA}] 
\label{eq:st1}
\end{equation}
\noindent
with
\begin{equation}
V[\vec{M}_q^{\tA}] = V[\bar{F}(\vec{Y}_\bullet,\vec{S}_q)]
+  \frac{1}{J} E[\hat{\bSig}_{\bar{M}_{j\bullet}^{\tT}} ]
\label{eq:int4}
\end{equation}
\noindent
and
\begin{equation}
Cov[\vec{M}_q^{\tA},\vec{M}_{q^\prime}^{\tA}] = \frac{1}{J} V[E[F(Y_j,S_q)|Y_j]] ,
\label{eq:st2}
\end{equation}
\noindent
in which case
\begin{equation}
E[\hat{\bSig}_{\vec{M}_q^{\tA}}] = V[\bar{F}(\vec{Y}_\bullet,\vec{S}_q)]
+  \frac{1}{J} \left( E[\hat{\bSig}_{\bar{M}_{j\bullet}^{\tT}} ] - V[E[F(Y_j,S_q)|Y_j]] \right) . \label{eq:int9}
\end{equation}
\noindent
Using Lemma 1, we write $E[\hat{\bSig}_{\bar{M}_{j\bullet}^{\tT}} ]$ in (\ref{eq:int4}) as
\begin{equation}
E[\hat{\bSig}_{\bar{M}_{j\bullet}^{\tT}} ] = V[\bar{M}_{j\bullet}^{\tT}]
-Cov[\bar{M}_{j\bullet}^{\tT},\bar{M}_{j^\prime\bullet}^{\tT}] .
\label{eq:st3}
\end{equation}
\noindent
Next,
\begin{eqnarray}
V[\bar{M}_{j\bullet}^{\tT}] \hspace*{-6mm} && = V \left[ \frac{1}{Q} \sum_{q=1}^Q F(Y_j,S_q)\right] \nonumber \\
&& = \frac{1}{Q^2} \sum_{q=1}^Q \sum_{q^\prime=1}^Q Cov[ F(Y_j,S_q), F(Y_j,S_{q^\prime})] \nonumber \\
&& = \frac{1}{Q} V[F(Y_j,S_q)] + \frac{Q-1}{Q} Cov[ F(Y_j,S_q), F(Y_j,S_{q^\prime})] .  \label{eq:int6}
\end{eqnarray}
\noindent
The $S_q$ in (\ref{eq:int6}) are independent and identically distributed so, applying Lemma 4,
$V[\bar{M}_{j\bullet}^{\tT}]$ in (\ref{eq:st3}) becomes, from (\ref{eq:int6}), 
\begin{equation}
V[\bar{M}_{j\bullet}^{\tT}] = \frac{1}{Q} V[F(Y_j,S_q)] + \frac{Q-1}{Q} V[E[F(Y_j,S_q)|Y_j]] . \label{eq:var2}
\end{equation}
\noindent
Similarly, the covariance $Cov[\bar{M}_{j\bullet}^{\tT},\bar{M}_{j^\prime\bullet}^{\tT}]$
in (\ref{eq:st3}) is 
\begin{eqnarray}
Cov[\bar{M}_{j\bullet}^{\tT},\bar{M}_{j^\prime\bullet}^{\tT}] \hspace*{-6mm}
&& = Cov\left[ \frac{1}{Q} \sum_{q=1}^Q F(Y_j,S_q) , \frac{1}{Q} \sum_{q^\prime=1}^Q F(Y_{j^\prime},S_{q^\prime})  \right]  \nonumber \\
&& = \frac{1}{Q^2} \sum_{q=1}^Q \sum_{q^\prime=1}^Q Cov[ F(Y_j,S_q), F(Y_{j^\prime},S_{q^\prime})] \nonumber \\
&& = \frac{1}{Q}  Cov[ F(Y_j,S_q), F(Y_{j^\prime},S_q)]  \nonumber \\
&& = \frac{1}{Q} E[ Cov[ F(Y_j,S_q), F(Y_{j^\prime},S_{q})|S_q]] \nonumber \\
&& \hspace*{10mm} +\, \frac{1}{Q} Cov[ E[F(Y_j,S_q)|S_q], E[F(Y_{j^\prime},S_q)|S_q]]  \nonumber \\
&& = \frac{1}{Q} V[E[F(Y_j,S_q)|S_q]]  . \label{eq:cov2}
\end{eqnarray}
Substituting (\ref{eq:var2}) and (\ref{eq:cov2}) back into (\ref{eq:st3}) yields
\begin{eqnarray}
E[\hat{\bSig}_{\bar{M}_{j\bullet}^{\tT}} ] \hspace*{-6mm}
&& = \frac{1}{Q} V[F(Y_j,S_q)] + \frac{Q-1}{Q} V[E[F(Y_j,S_q)|Y_j]] \nonumber \\[-2mm]
&&   \label{eq:resES2} \\[-2mm]
&& \hspace*{10mm} -\,\frac{1}{Q} V[E[F(Y_j,S_q)|S_q]] .  \nonumber
\end{eqnarray}
\noindent
Finally, substituting this back into (\ref{eq:int9}) proves (\ref{eq:th22}). \hfill $\qed$

\vspace*{2mm}
For the scalar case $K=1$ the relative bias
\begin{equation}
\mbox{relbias}[ \hat{\bSig}_{\vec{M}_q^{\tA}} ]
= \frac{1}{JQ}\frac{\Phi}{V[\bar{F}(\vec{Y}_\bullet,\vec{S}_q)]}
\label{eq:relbias2}
\end{equation}
\noindent
associated with the MC sample variance in Proposition 2 has simple bounds, given in the following
proposition. Following the proof of this proposition, we show by simple examples that these bounds are tight.

%--------------------------------------------------------------------------------------------------------------- Proposition 3

\vspace*{2mm}
{\it Proposition 3}: The relative bias in (\ref{eq:relbias2}) satisfies
\begin{equation}
0 \leq \mbox{relbias}[ \hat{\bSig}_{\vec{M}_q^{\tA}} ] \leq \frac{1}{Q} .
\label{eq:prop3}
\end{equation}

%--------------------------------------------------------------------------------------------------------------- Proof
\vspace*{2mm}
{\it Proof}: We prove first that the relative bias in (\ref{eq:prop3}) is non-negative. The variance of a
random variable $X$ can be expressed by
\begin{equation}
V[X]=\frac{1}{2}E[(X-X^\prime)^2]
\label{eq:altvar}
\end{equation}
\noindent
where $X,X^\prime$ are independent and identically distributed. Using conditional versions of (\ref{eq:altvar}),
we have
\begin{eqnarray}
\Phi \hspace*{-6mm}
&& = E[V[F(Y_j,S_q)|S_q] ] - V[E[F(Y_j,S_q)|Y_j] ]                              \nonumber \\
&& = E\left[ \frac{1}{2} E[(F(Y_j,S_q)-F(Y_j^\prime,S_q))^2|S_q] \right]     \nonumber  \\
&& \hspace*{10mm} -\,\frac{1}{2}E\left[ (E[F(Y_j,S_q)|Y_j]-E[F(Y_j^\prime,S_q)|Y_j^\prime])^2 \right]  . \nonumber
\end{eqnarray}
\noindent
Define $\Delta(y,y^\prime,s) = F(y,s)-F(y^\prime,s)$. Then
\begin{eqnarray}
\Phi \hspace*{-6mm}
&& = \frac{1}{2} E[\Delta^2(Y_j,Y_j^\prime,S_q)]  
-  \frac{1}{2}E\left[ E^2[\Delta(Y_j,Y_j^\prime,S_q)|Y_j,Y_j^\prime] \right]   \nonumber  \\
&& = \frac{1}{2} E\left[ E[\Delta^2(Y_j,Y_j^\prime,S_q)|Y_j,Y_j^\prime] \right]  
-  \frac{1}{2}E\left[ E^2[\Delta(Y_j,Y_j^\prime,S_q)|Y_j,Y_j^\prime] \right]   \nonumber   \\
&& = \frac{1}{2} E\left[ V[\Delta(Y_j,Y_j^\prime,S_q)|Y_j,Y_j^\prime] \right]  . \nonumber 
\end{eqnarray}
\noindent
Variance is non-negative so $\Phi \geq 0$, proving the lower bound in Proposition 3. To prove the upper bound
we note first that, applying Lemma 4 to the scalar case $V[\bar{F}(Y_\bullet,S_q)]$ of the target variance in
(\ref{eq:simp2}), we have
\begin{eqnarray}
V[\bar{F}(Y_\bullet,S_q)] \hspace*{-6mm}
&& = \frac{1}{J} V[F(Y_j,S_q)] + \frac{J-1}{J} V[E[F(Y_j,S_q) | S_q]   \nonumber \\
&& = \frac{1}{J} E[V[F(Y_j,S_q) | S_q] + V[E[F(Y_j,S_q) | S_q]  . \nonumber
\end{eqnarray}
\noindent
Then the relative bias given in (\ref{eq:relbias2}) is
\begin{eqnarray}
\mbox{relbias}[ \hat{\bSig}_{\vec{M}_q^{\tA}} ] \hspace*{-6mm}
&& = \frac{1}{JQ}\frac{E[V[F(Y_j,S_q)|S_q] - V[E[F(Y_j,S_q)|Y_j]}{\frac{1}{J} E[V[F(Y_j,S_q) | S_q] + V[E[F(Y_j,S_q) | S_q]}
\nonumber \\
&& \leq \frac{1}{JQ}\frac{E[V[F(Y_j,S_q)|S_q]}{\frac{1}{J} E[V[F(Y_j,S_q) | S_q]} \nonumber \\
&& = \frac{1}{Q} ,
\end{eqnarray}
\noindent
establishing the upper bound in (\ref{eq:prop3}). \hfill $\qed$

\vspace*{2mm}
In the remainder of this section we look at the examples of additive and multiplicative error to see that the bounds
in Proposition 3 on the relative bias of the MC sample variance $\hat{\bSig}_{\vec{M}_q^{\tA}}$ are tight.

\vspace*{2mm}
{\it Additive error}: In this case $F(\vec{y},\vec{s}) = \vec{y}+\vec{s}$ and $\Phi$ in (\ref{eq:Phi}) is
\begin{eqnarray}
\Phi \hspace*{-6mm}
&&= E[V[\vec{Y}_j+\vec{S}_q|\vec{S}_q]] - V[E[\vec{Y}_j+\vec{S}_q|\vec{Y}_j]] \nonumber \\
&&= E[V[\vec{Y}_j]] - V[\vec{Y}_j+E[\vec{S}_q]] \nonumber \\
&&= 0 .
\end{eqnarray}
\noindent
This shows that Proposition 3's lower bound is tight for the scalar case addressed there. More generally,
it shows for additive shared systematic error that {\sf Combine} MC replicates constructed according to
(\ref{eq:comM2}) have both zero mean bias and zero covariance bias.

\vspace*{2mm}
{\it Multiplicative error}: We focus on the univariate case $K=1$ in which we have scalars, $Y_j$ and $S_q$. Then
\begin{eqnarray}
\Phi \hspace*{-6mm}
&&= E[V[Y_j S_q|S_q]] - V[E[Y_j S_q|Y_j]] \nonumber \\
&&= E[S_q^2]\,V[Y_j] - E^2[S_q]\,V[Y_j] \nonumber \\
&&= V[S_q]\,V[Y_j] .
\end{eqnarray}
\noindent
The corresponding relative bias for our case $K=1$ is
\begin{eqnarray}
\mbox{relbias}[ \hat{\bSig}_{\vec{M}_q^{\tA}} ] \hspace*{-6mm}
&&= \frac{1}{JQ}\frac{\Phi}{V[\bar{F}(Y_\bullet,{S_q})]}  \nonumber \\
&&= \frac{1}{Q} \frac{V[S_q]V[\bar{Y}_\bullet]}{V[\bar{Y}_\bullet{S_q}]}  \nonumber \\
&&= \frac{1}{Q}
\frac{V[S_q]V[\bar{Y}_\bullet]}{V[S_q]V[\bar{Y}_\bullet]+V[S_q]E^2[Y_j]+V[\bar{Y}_\bullet]E^2[S_q]} 
\label{eq:bound}
\end{eqnarray} 
\noindent
where in the last step we used the product rule for variance \cite{goodman}. Expression (\ref{eq:bound})
is less than or equal to $1/Q$, achieving $1/Q$ for $E[Y_j]=E[S_q]=0$, showing that the
upper bound in Proposition 3 is tight. Thus, for multiplicative error with the alternative MC sample
construction, the relative covariance bias can be as great as 100\% in the extreme case $Q=1$; but
for typical user choices of $Q$ it is no greater than a small fraction of a percent.

\vspace*{2mm}
We conducted a computer experiment as a numerical check on the upper bound in Proposition 3 and, in particular,
on (\ref{eq:bound}) for relative bias in the case of multiplicative error. {\sf Combine} MC samples with sizes ranging
from $Q=3$ to $Q=300$ were constructed per the proposed alternative method for the univariate case ($K=1$)
with independent standard normal datasets $Y_j$ of size $J=4$ and independent standard normal errors $S_q$.
For each sample size 10,000 MC samples were created and their 10,000 sample variances were averaged.
These averaged sample variances were compared against the target variance
\begin{displaymath}
V[\bar{F}(\vec{Y}_\bullet,\vec{S}_q)] = V[S_q]V[\bar{Y}_\bullet] = V[S_q]\frac{V[Y_j]}{J} = 1/4
\end{displaymath}
\noindent
to estimate for each $Q$ the relative bias in the MC sample variance. The estimated relative biases are plotted
in Fig.\ 4. According to (\ref{eq:bound}), the estimated relative biases should agree with the solid red line in Fig.\ 4
given by $\mbox{relbias} = 1/Q$. They do agree to within experimental uncertainty.  

\begin{figure}
\vspace*{-20mm}
\begin{center}
\includegraphics[width=4.4in,height=3in]{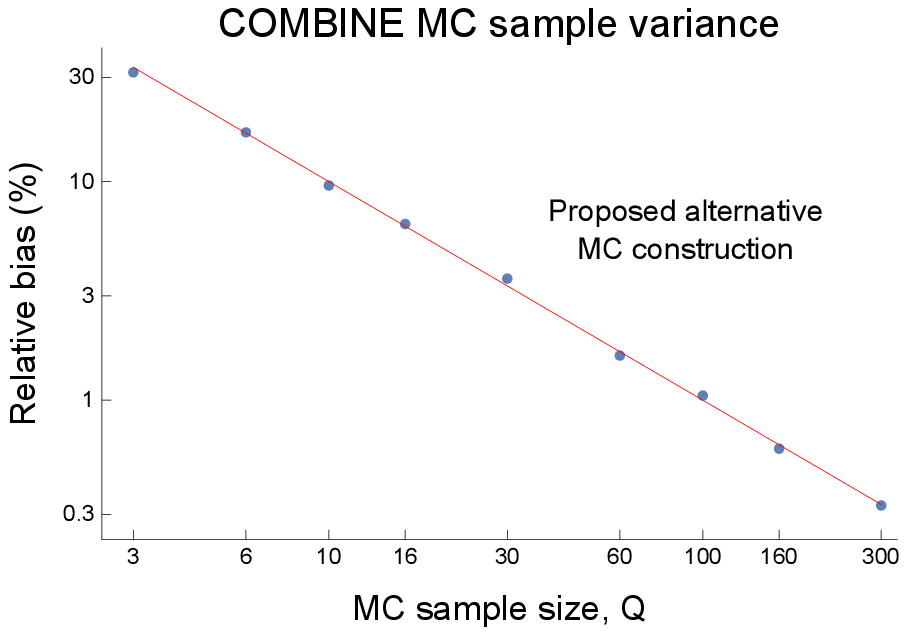}
\end{center}
\begin{center}
\parbox{3.9in}{\small Figure 4. Results of an experiment confirming the tightness of the upper bound $1/Q$ on the relative bias
of the sample variance with the proposed alternative construction of {\sf Combine} MC replicates.}
\end{center}
\vspace*{2mm}
\end{figure}

\section{Relative variability}

Sections 2 and 3 compared {\sf Combine}'s current and proposed alternative constructions of MC replicates from the
standpoint of bias in the MC sample means and covariances. Specifically, Propositions 1 and 2 established that the MC sample
means $\bar{M}_\bullet^{\tC}$ and $\bar{M}_\bullet^{\tA}$ are each unbiased. These propositions also established that
MC sample covariance $\hat\bSig_{\vec{M}_q^{\tC}}^2$ is biased, while the bias of $\hat\bSig_{\vec{M}_q^{\tA}}^2$ is
asymptotically zero as $Q\rightarrow\infty$. In this section we complete our comparison of the two constructions
by considering the differences in the variabilities of their MC sample means and covariances.

\subsection{Relative variability in MC sample means}

The MC sample means with the current and alternative MC constructions (\ref{eq:comM}) and (\ref{eq:comM2})
are, respectively,
\begin{eqnarray}
\bar{M}_\bullet^{\tC} \hspace*{-6mm} && = \bar{M}_{\bullet\bullet}^{\tT}
+ \frac{1}{\sqrt{J}} \bUc \sqrt{\bDc} \bar{Z}_\bullet         ,      \nonumber \\[-3mm]
\label{eq:constructs}                                                                             \\[-1mm]
\bar{M}_\bullet^{\tA} \hspace*{-6mm} && = \bar{M}_{\bullet\bullet}^{\tT}
+ \frac{1}{\sqrt{J}} \bUa \sqrt{\bDa} \bar{Z}_\bullet  .      \nonumber
\end{eqnarray}
\noindent
The difference in their degrees of variability (their covariances) is given by the following proposition.

%--------------------------------------------------------------------------------------------------------------- Proposition 4
\vspace*{2mm}
{\it Proposition 4}: Consider the two-stage scenario in Fig.\ 1 with the same set-up as in Propositions 1 and 2. Then
\begin{equation}
V[\bar{M}_\bullet^{\tC}] - V[\bar{M}_\bullet^{\tA}] = \frac{1}{JQ} \Psi - \frac{1}{JQ^2} \Phi
\label{eq:diff}
\end{equation}
\noindent
with the matrices $\Psi$ and $\Phi$ as defined in (\ref{eq:Psi}) and (\ref{eq:Phi}).

\vspace*{2mm}
%--------------------------------------------------------------------------------------------------------------- Proof
{\it Proof}: Lemma 2 yields for the MC sample means in (\ref{eq:constructs}) that
\begin{eqnarray}
V[\bar{M}_\bullet^{\tC}] \hspace*{-6mm} && = V[\bar{M}_{\bullet\bullet}^{\tT}]
+ \frac{1}{JQ} E[\hat\bSig_{\vec{N}_j^{\tT}}]  ,            \nonumber  \\
V[\bar{M}_\bullet^{\tA}] \hspace*{-6mm} && = V[\bar{M}_{\bullet\bullet}^{\tT}]
+ \frac{1}{JQ} E[\hat\bSig_{\bar{M}_{j\bullet}^{\tT}}]    ,      \nonumber
\end{eqnarray}
\noindent
in which case
\begin{equation}
V[\bar{M}_\bullet^{\tC}] - V[\bar{M}_\bullet^{\tA}] = \frac{1}{JQ}
E[\hat\bSig_{\vec{N}_j^{\tT}}] - \frac{1}{JQ} E[\hat\bSig_{\bar{M}_{j\bullet}^{\tT}}]  .
\nonumber
\end{equation}
\noindent
Using the definitions of $\Psi$ and $\Phi$ and the results in (\ref{eq:resES1}) and (\ref{eq:resES2})
for $E[\hat\bSig_{\vec{N}_j^{\tT}}]$ and $E[\hat\bSig_{\bar{M}_{j\bullet}^{\tT}}]$, we find
\begin{eqnarray}
V[\bar{M}_\bullet^{\tC}] - V[\bar{M}_\bullet^{\tA}]  && \nonumber \\
&& \hspace*{-25mm} = \frac{1}{JQ}V[F(\vec{Y}_j,\vec{\nu})]
- \frac{1}{JQ} \left( \frac{1}{Q} V[F(\vec{Y}_j,\vec{S}_q)] \right.  \nonumber  \\
&& \hspace*{-15mm} -\, \frac{1}{Q} \left. V[E[F(\vec{Y}_j,\vec{S}_q)|\vec{S}_q]]
+\frac{Q-1}{Q} V[E[F(\vec{Y}_j,\vec{S}_q)|\vec{Y}_j]]       
 \right)            \nonumber \\
&& \hspace*{-25mm} = \frac{1}{JQ} \Psi - \frac{1}{JQ} \left(  \frac{1}{Q} V[F(\vec{Y}_j,\vec{S}_q)] \right.  \nonumber  \\
&& \hspace*{-15mm} -\,    \frac{1}{Q} \left. V[E[F(\vec{Y}_j,\vec{S}_q)|\vec{S}_q]]    
- \frac{1}{Q} V[E[F(\vec{Y}_j,\vec{S}_q)|\vec{Y}_j]] \right)            \nonumber \\
&& \hspace*{-25mm} = \frac{1}{JQ} \Psi - \frac{1}{JQ^2} \left(  E[V[F(\vec{Y}_j,\vec{S}_q)|\vec{S}_q]]
- V[E[F(\vec{Y}_j,\vec{S}_q)|\vec{Y}_j]]  \right)  .         \nonumber
\end{eqnarray}
\noindent
Recalling definition (\ref{eq:Phi}) for $\Phi$, this proves the proposition. \hfill $\qed$

\vspace*{2mm}
For sufficiently large finite $Q$, the sign of the difference in the covariances $V[\bar{M}_\bullet^{\tC}]$ and
$V[\bar{M}_\bullet^{\tA}]$ is determined by $\Psi$ in (\ref{eq:diff}). We saw in Sect.\ 2 that $\Psi$ can
be positive, negative, or zero, so depending on the form of the {\sf Transform} error model $F(\vec{Y}_j,\vec{S}_q)$
either of the two estimators $\bar{M}_\bullet^{\tC}$ or $\bar{M}_\bullet^{\tA}$ can exhibit less variability.
According to (\ref{eq:diff}), when $\Psi$ is positive, the
alternatively constructed MC replicates $\vec{M}_q^{\tA}$ are better than the current $\vec{M}_q^{\tC}$ both
because their sample mean $\bar{M}_\bullet^{\tA}$ exhibits less variability and because their sample variance
$\hat{\bSig}_{\bar{M}_{j\bullet}^{\tA}}^2$ is asymptotically unbiased. When $\Psi$ is negative, the picture is
mixed: $\vec{M}_q^{\tA}$ exhibits greater variability than $\vec{M}_q^{\tC}$, but $\vec{M}_q^{\tC}$ is biased.
All these considerations of the relative variabilities of $\bar{M}_\bullet^{\tC}$ and $\bar{M}_\bullet^{\tA}$ are,
of course, dominated by Proposition 4's main import that the difference in their variabilties is asymptotically zero,
\begin{equation}
V[\bar{M}_\bullet^{\tC}] - V[\bar{M}_\bullet^{\tA}] = 0 \hspace*{2mm} \mbox{for }  Q\rightarrow\infty .
\nonumber
\end{equation}

\subsection{Relative variability in MC sample covariances}

We consider in this subsection the difference in the variabilities of the MC sample covariances with the current
and proposed alternative constructions, limiting our considerations to the univariate ($K=1$) case. For $K=1$
\begin{equation}
M_q^{\tC} = \bar{M}_{\bullet q}^{\tT} + \frac{Z_q}{\sqrt{J}} \sqrt{S_{N_j^{\tT}}^2}
\label{eq:univ}
\end{equation}
\noindent
where $N_j^{\tT} = F(Y_j,\nu)$ and
\begin{displaymath}
S_{N_j^{\tT}}^2
= \frac{1}{J-1} \sum_{j=1}^J\left(N_j^{\tT} - \bar{N}_\bullet^{\tT}\right)^2.
\end{displaymath}
\noindent
Even with the restriction $K=1$, assessing the variance of the sample variance of the $M_q^{\tC}$ in (\ref{eq:univ})
is difficult, necessarily involving fourth moments. Because the $Z_q$ in (\ref{eq:univ}) are normal, the following
lemma (proved in \cite{knautz}) is useful. 

\vspace*{2mm}
{\it Lemma 5}: Let $X_n = \mu_n +\sigma Z_n$ for $n=1, ..., N$ where $\sigma$ and the $\mu_n$ are constants
and the $Z_n$ are mutually independent and standard normal-distributed. Let
\begin{equation}
u^2 = \frac{1}{N-1}\sum_{n=1}^N (\mu_n-\bar{\mu})^2 , \;\;\;
S_X^2 = \frac{1}{N-1}\sum_{n=1}^N (X_n-\bar{X})^2 .
\nonumber
\end{equation}
\noindent
Then $E[S_X^2] = u^2 +\sigma^2$ and
\begin{equation}
V[S_X^2] = \frac{2}{N-1} \sigma^4 + \frac{4}{N-1} \sigma^2 u^2 .
\end{equation}

\vspace*{2mm}
Let $\bar{M}_{\bullet}^{\tT}$ be the set of $Q$ MC sample means $\bar{M}_{\bullet q}^{\tT}$.
Conditioned on $\bar{M}_{\bullet}^{\tT}$ and $S_{N_j^{\tT}}^2$, {\sf Combine}'s MC
replicates $M_q^{\tC}$ in (\ref{eq:univ}) are independent and normally distributed with means
$\bar{M}_{\bullet q}^{\tT}$ and variance $\frac{1}{J}S_{N_j^{\tT}}^2$. Then according to
Lemma 5,
\begin{equation}
E[S_{M_q^{\tC}}^2|\bar{M}_{\bullet}^{\tT},S_{N_j^{\tT}}^2]
= \frac{1}{J}S_{N_j^{\tT}}^2 + S_{\bar{M}_{\bullet q}^{\tT}}^2
\nonumber
\end{equation}
\noindent
and
\begin{equation}
V[S_{M_q^{\tC}}^2|\bar{M}_{\bullet}^{\tT},S_{N_j^{\tT}}^2]
= \frac{2}{Q-1}\frac{1}{J^2}S_{N_j^{\tT}}^4
+ \frac{4}{Q-1} \frac{1}{J} S_{N_j^{\tT}}^2 S_{\bar{M}_{\bullet q}^{\tT}}^2
\nonumber
\end{equation}
\noindent
where
\begin{equation}
S_{\bar{M}_{\bullet q}^{\tT}}^2 = \frac{1}{Q-1} \sum_{q=1}^Q
\left( \bar{M}_{\bullet q}^{\tT} - \bar{M}_{\bullet \bullet}^{\tT} \right)^2 .
\nonumber
\end{equation}
\noindent
Then
\begin{eqnarray}
V[S_{M_q^{\tC}}^2] \hspace*{-6mm}
&& =V[E[S_{M_q^{\tC}}^2|\bar{M}_{\bullet}^{\tT},S_{N_j^{\tT}}^2]] \nonumber \\
&& \hspace*{10mm} +\,E[V[S_{M_q^{\tC}}^2|\bar{M}_{\bullet}^{\tT},S_{N_j^{\tT}}^2]]  \nonumber \\
&& = V\left[   \frac{1}{J}S_{N_j^{\tT}}^2 + S_{\bar{M}_{\bullet q}^{\tT}}^2  \right]   \nonumber \\
&& \hspace*{10mm} +\,\frac{2}{Q-1}\frac{1}{J^2} E\left[   S_{N_j^{\tT}}^4\right]
+ \frac{4}{Q-1} \frac{1}{J} E\left [S_{N_j^{\tT}}^2 S_{\bar{M}_{\bullet q}^{\tT}}^2   \right] .
\label{eq:vc}
\end{eqnarray}
\noindent
A parallel calculation for the alternatively constructed MC replicates
\begin{displaymath}
M_q^{\tA} = \bar{M}_{\bullet q}^{\tT} + \frac{Z_q}{\sqrt{J}} \sqrt{S_{\bar{M}_{j\bullet}^{\tT}}^2}
\end{displaymath}
\noindent
yields
\begin{eqnarray}
V[S_{M_q^{\tA}}^2] \hspace*{-6mm}
&& = V\left[   \frac{1}{J}S_{\bar{M}_{j\bullet}^{\tT}}^2 + S_{\bar{M}_{\bullet q}^{\tT}}^2  \right] \nonumber \\
&& \hspace*{10mm} +\,\frac{2}{Q-1}\frac{1}{J^2} E\left[   S_{\bar{M}_{j\bullet}^{\tT}}^4\right]
+ \frac{4}{Q-1} \frac{1}{J} E\left [S_{\bar{M}_{j\bullet}^{\tT}}^2 S_{\bar{M}_{\bullet q}^{\tT}}^2   \right] .
\label{eq:va}
\end{eqnarray}
\noindent
We have, therefore, from (\ref{eq:vc}) and (\ref{eq:va}) that
\begin{eqnarray}
V[S_{M_q^{\tA}}^2] -V[S_{M_q^{\tC}}^2]  &&    \nonumber \\
&& \hspace*{-30mm} \approx  V\left[   \frac{1}{J}S_{\bar{M}_{j\bullet}^{\tT}}^2 + S_{\bar{M}_{\bullet q}^{\tT}}^2  \right]
- V\left[   \frac{1}{J}S_{N_j^{\tT}}^2 + S_{\bar{M}_{\bullet q}^{\tT}}^2  \right] + O(1/Q) 
\nonumber \\
&& \hspace*{-30mm} \approx  \frac{1}{J^2} \left(  V[  S_{\bar{M}_{j\bullet}^{\tT}}^2]
-V[  S_{N_j^{\tT}}^2   ] \right) + O(1/Q) .
\label{eq:vardiff}
\end{eqnarray}
\noindent
We now pursue expressions for the difference in the two variances in (\ref{eq:vardiff}) in the
cases of additive and multiplicative error.

\vspace*{2mm}
{\it Additive error model}: For $f(y,s) =y+s$ we have $\bar{M}_{j\bullet}^{\tT} = Y_j+\bar{S}_{\bullet}$ and
$N_j^{\tT}=Y_j+\nu$ where $\nu = E[S_q]$. Then $S_{\bar{M}_{j\bullet}^{\tT}}^2 = S_{Y_j}^2$ and
$S_{N_j^{\tT}}^2 = S_{Y_j}^2$ so
\begin{equation}
V[  S_{\bar{M}_{j\bullet}^{\tT}}^2] - V[S_{N_j^{\tT}}^2] = 0 .
\label{eq:additiv}
\end{equation}
 
\vspace*{2mm}
{\it Multiplicative error model}: For $f(y,s) = ys$ we have $\bar{M}_{j\bullet}^{\tT} = Y_j\bar{S}_{\bullet}$
and $N_j^{\tT}=Y_j\nu$. Then $S_{\bar{M}_{j\bullet}^{\tT}}^2 = \bar{S}_{\bullet}^2 S_{Y_j}^2$ and
$S_{N_j^{\tT}}^2 = \nu^2 S_{Y_j}^2$. The factors $\bar{S}_{\bullet}^2$ and $S_{Y_j}^2$ are independent,
so using the product rule for variance \cite{goodman}, we find
\begin{eqnarray}
V[S_{\bar{M}_{j\bullet}^{\tT}}^2] - V[S_{N_j^{\tT}}^2]  \hspace*{-6mm}
&& = V[\bar{S}_{\bullet}^2 S_{Y_j}^2] - V[  \nu^2 S_{Y_j}^2]     \nonumber \\
&& = V[\bar{S}_\bullet^2]E^2[S_{Y_j}^2] + E[\bar{S}_{\bullet}^4] V[S_{Y_j}^2] - \nu^4 V[S_{Y_j}^2] . 
\label{eq:multi1}
\end{eqnarray}
\noindent
Let $\tau$, $\omega$, and $\psi$ be the second, third, and fourth central moments of $S_q$, and let $\phi$ be the fourth central
moment of $Y_j$. Moment results for the sample mean and sample variance in \cite{angelova, rose} then yield
\begin{eqnarray}
E[S_{Y_j}^2] \hspace*{-6mm} && = \sigma^2 ,   \nonumber \\
V[S_{Y_j}^2] \hspace*{-6mm} && = \frac{\phi^4-\sigma^4}{J} + \frac{2\sigma^4}{J(J-1)}, \nonumber \\
E[\bar{S}_{\bullet}^4] \hspace*{-6mm} && =  \nu^4 + \frac{6\nu^2\tau^2}{Q}
+ \frac{3\tau^4+4\nu\omega^3}{Q^2} + \frac{\psi^4-3\tau^4}{Q^3},   \nonumber \\
V[\bar{S}_{\bullet}^2] \hspace*{-6mm}  &&   =  \frac{4\nu^2\tau^2}{Q} 
+ \frac{2\tau^4+4\nu\omega^3}{Q^2} + \frac{\psi^4-3\tau^4}{Q^3}. \nonumber
\end{eqnarray}
\noindent 
It then follows from (\ref{eq:multi1}) that
\begin{eqnarray}
V[S_{\bar{M}_{j\bullet}^{\tT}}^2] - V[S_{N_j^{\tT}}^2]  \hspace*{-6mm}
&& = \sigma^4\!\left( \frac{4\nu^2\tau^2}{Q}\!+\!\frac{2\tau^4+4\nu\omega^3}{Q^2}\!+\!\frac{\psi^4-3\tau^4}{Q^3}
\right)    \nonumber \\[-3mm]
\label{eq:diffv} \\[-1mm]
&& \hspace*{-30mm}+\,\left( \frac{6\nu^2\tau^2}{Q}\!+\!\frac{3\tau^4+4\nu\omega^3}{Q^2}\!+\!\frac{\psi^4-3\tau^4}{Q^3}
\right)\!\!\left( \frac{\phi^4-\sigma^4}{J}\!+\!\frac{2\sigma^4}{J(J-1)}
\right) . \nonumber
\end{eqnarray}
\noindent
This suggests generally for multiplicative error that, for $Q\rightarrow\infty$,
\begin{equation}
V[S_{M_q^{\tA}}^2] -V[S_{M_q^{\tC}}^2] \rightarrow 0 .
\label{eq:multiplicativ}
\end{equation}

\vspace*{2mm}
Computer experiments were done to check the approximation in (\ref{eq:vardiff}) for the difference in the variances of the MC sample
variances. According to results (\ref{eq:additiv}) and (\ref{eq:multiplicativ}), the difference in the variances should be zero and
asymptotically zero for additive and multiplicative noise, respectively. These results were confirmed in experiments with different
distributions for $Y_j$ and $S_q$ and different  data sample sizes $J$. The results in Fig.\ 5 are typical. The plots in Fig.\ 5 show
the estimated relative difference
\begin{equation}
\mbox{reldiff} = \frac{\hat{V}[S_{M_q^{\tC}}^2] -\hat{V}[S_{M_q^{\tA}}^2]}
{\hat{V}[S_{M_q^{\tC}}^2]+\hat{V}[S_{M_q^{\tA}}^2]} ,
\label{eq:def}
\end{equation}
\noindent
the top plot for a case of additive noise and the bottom plot for a case of multiplicative noise. In each plot the
data and noise are standard normal-distributed and $J=4$. The plotted relative differences defined by
(\ref{eq:def}) have a potential range of $\pm100\%$. Each plotted point in Fig.\ 5 is estimated from an independent
set of 100,000 computer trials. The experiment results in Fig.\ 5 and corresponding results obtained for other
distributions and sample sizes confirm (\ref{eq:additiv}) and (\ref{eq:multiplicativ}) under broad conditions. 

\begin{figure}[b!]
\vspace*{8mm}
\begin{center}
\includegraphics[width=3.8in,height=4.5in]{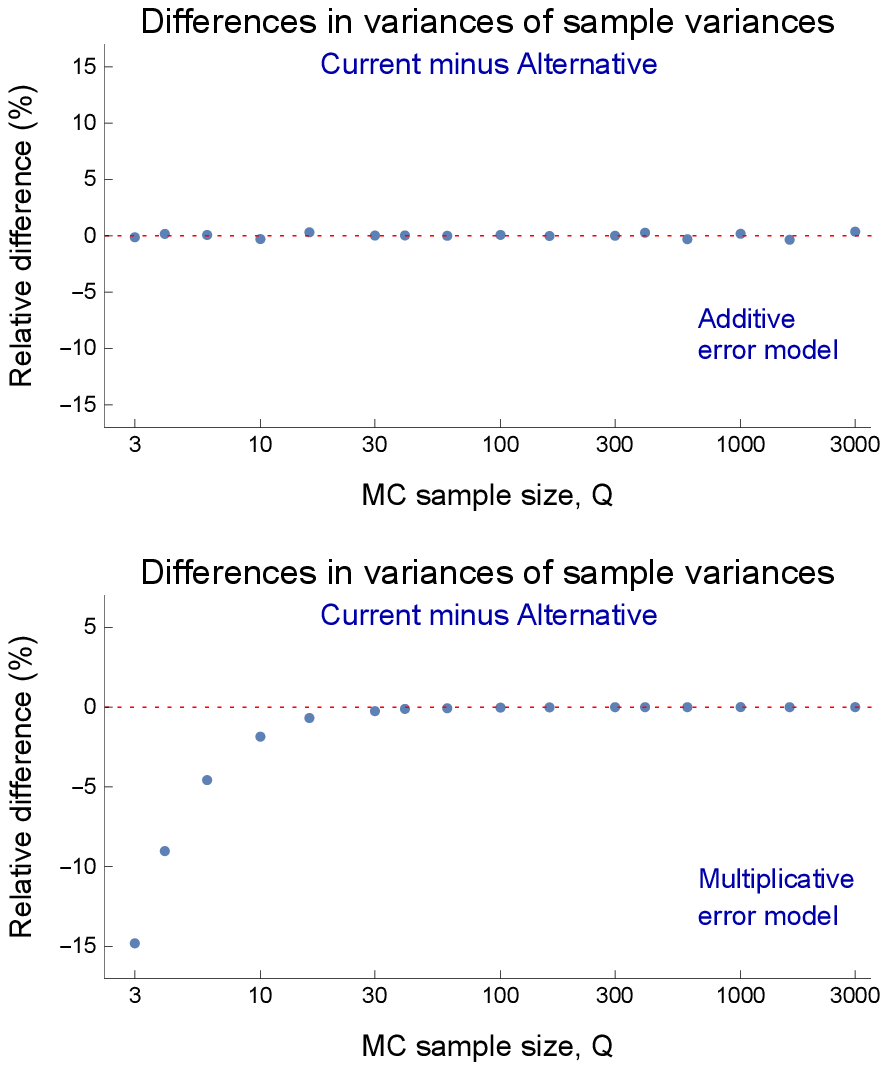}
\end{center}
\vspace*{-1mm}
\begin{center}
\parbox{4.25in}{\small Figure 5. Simulation-estimated relative differences in the variances of the MC sample variances
based on the current and proposed alternative methods, for additive error (top) and multiplicative error (bottom).}
\end{center}
\end{figure}

\vspace*{2mm}
Our aim with results (\ref{eq:additiv}) and (\ref{eq:multiplicativ}) was to discover whether either method of constucting
MC replicates dominates the other with respect to variance of the MC sample variance. These results, and the experiments
confirming them, indicate that neither MC replicate construction method dominates the other when the error is additive or
multiplicative. For error models beyond the additive and multiplicative cases, little analytical headway seems possible,
so we turn directly to computer experiments.

\vspace*{2mm}
Presented in Fig.\ 6 are experiment results obtained for the phase and exponential error models. The top plot
in Fig.\ 6 shows the relative difference (\ref{eq:def}) of variances for phase error for the extremal case discussed in Sect.\ 2
in which the error $S_q$ is distributed $\mbox{Unif}[-\pi,\pi]$, the data $Y_j$ are $-\pi/2$ and $\pi/2$ with equal
probabilities, and $J=4$. The middle and bottom plots show two cases of exponential error, both in which the data and
exponential error are uniformly distributed, $Y_j \sim \mbox{Unif}[0,b]$ and $S_q \sim \mbox{Unif}[1-\alpha,1+\alpha]$,
with $b=8$ and $\alpha = .95$ in the middle plot and $b=1$ and $\alpha = .95$ in the bottom plot. These two cases
of exponential error are the two cases in Fig.\  3 where the relative bias of {\sf Combine}'s MC sample variance is most
extreme---approaching $\pm$20\%.

\vspace*{2mm}
The results in Fig.\ 6 show that neither MC construction method dominates the other by having consistently smaller
variance in its sample variance. In the middle plot the relative difference (\ref{eq:def}) in variances is negative, meaning
that the sample variance with the current method has less variability. In the bottom plot, though, also with the exponential
error model, the sample variance with the alternative method has less variability. Also, the three plots illustrate that the relative
difference in variances can exhibit different degrees and types of transient behavior for small $Q$.
The bottom plot shows almost no transient change, while the middle plot shows significant transient change before settling
toward a limiting non-zero relative difference. The top plot shows that the relative difference in the
variances can even change sign before approaching its limit value.

\begin{figure}[h!]
\vspace*{-9mm}
\begin{center}
\includegraphics[width=3.6in,height=6.6in]{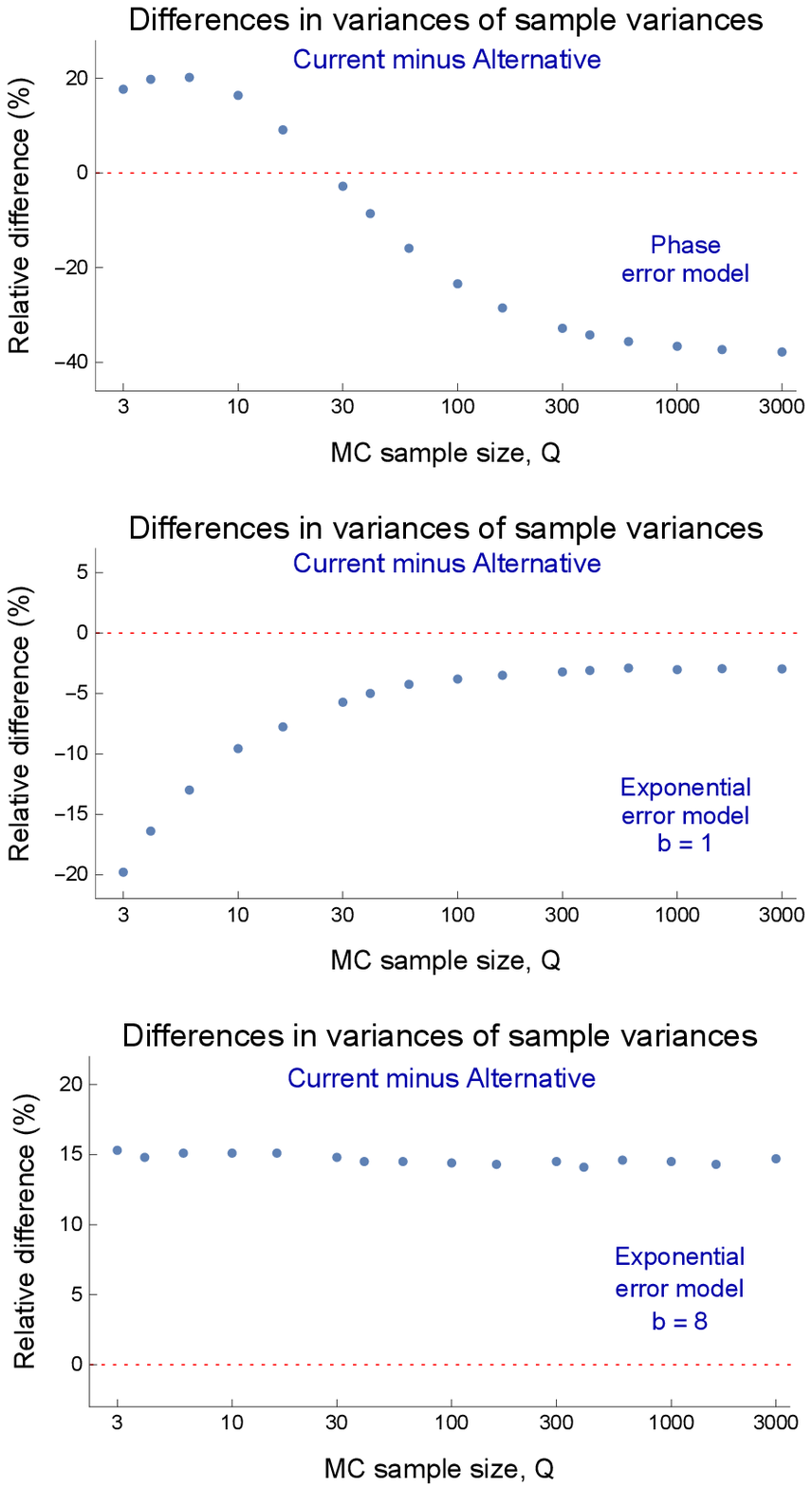}
\end{center}
\vspace*{-2mm}
\begin{center}
\parbox{4.4in}{\small Figure 6. Simulation-estimated relative differences in the variances of the MC sample variances
based on the current and proposed alternative methods, for phase error (top) and exponential error (middle and bottom).}
\end{center}
\end{figure}

\section{Summary remarks}

The MUF is a powerful tool for uncertainty modeling and analysis relating to data obtained in high-precision microwave
experiments, and the {\sf Combine} module is a key component of the MUF. We compared the MC replicates currently
constructed by {\sf Combine} with those based on an alternative construction, using bias and variance of the MC sample
mean and sample covariance as performance measures. We showed first that, with the current method of {\sf Combine}
MC replicate construction, the MC sample covariance is biased. Examples showed that this bias can be unacceptably
large---200\% in one extreme example and approaching $\pm$20\% in others---and cannot be reduced to a tolerable
level by choosing the MC sample size $Q$ sufficiently large. The MC sample covariance using the alternative construction
for MC replicates is also biased, but this bias is asymptotically zero with $Q$.

\vspace*{2mm}
Bias is the primary concern in MC sampling and the distinction, the current method being biased and the
alternative being asymptotically unbiased, is the two construction methods' most important difference.
Looking beyond bias to the difference in the variabilities of the MC sample means with the two methods, we
showed that this difference is asymptotically zero, with neither method dominating the other for small $Q$.
Comparing the variabilities of the MC sample variances was similarly nuanced and non-determinative: in the
cases of additive and multiplicative error, the difference in the variances of the sample variances is zero or
asymptotically zero, while for phase and exponential error, neither method consistently out-performs the other.

\vspace*{2mm}
We showed in this case study of bias in uncertainty propogation software that our proposed alternative MC replicate construction
method has an important advantage with regard to MC sample covariance bias, while lacking any clear disadvantage relative to
the current method. Consequently, the current method of constructing MC replicates in the MUF {\sf Combine} module is set to
be replaced with our proposed alternative. This study shows both that unknown, inadvertent biases are potentially present in even
well-designed statistical software and that statistical experiments can successfully identify these biases. We urge that statistical performance
tests be standard for modern software uncertainty propagation tools, and we anticipate that the statistical approach used here will be
useful to future analyses of MUF performance and of the performance of other similar statistical software for uncertainty propagation.

\vspace*{4mm}
\noindent
{\bf\Large Acknowledgment}

\vspace*{2mm}
This investigation benefited from early-stage discussions with Chih-Ming Wang and Sarah Streett, members
of the Statistical Engineering Division of the National Institute of Standards and Technology.

\noindent
{\bf\Large Appendix}

\vspace*{2mm}
We prove the four lemmas used in the proof of Proposition 1.

%--------------------------------------------------------------------------------------------------------------- Lemma 1
\vspace*{2mm}
{\it Lemma 1}: Let $\vec{X}_j\sim (\vec{\mu},\bSig)$ for $j=1, ...,J$, $J>1$ with common cross-covariance
$Cov[\vec{X}_j,\vec{X}_k]=\bSig^{\prime}$ for all $j\neq k$. Then $E[\hat{\bSig}]=\bSig-\bSig^{\prime}$
where $\hat{\bSig}$ is the sample covariance matrix
\begin{equation}
\hat{\bSig} = \frac{1}{J-1}\sum_{j=1}^J \left( \vec{X}_j-\bar{X}\right)\left(\vec{X}_j-\bar{X} \right)^\sT. 
\end{equation}

\vspace*{2mm}
{\it Proof}: 
\begin{eqnarray}
\hspace*{6mm} E[\hat{\bSig}] \hspace*{-6mm}
&& = E\left[ \frac{1}{J-1}\sum_{j=1}^J \vec{X}_j \vec{X}_j^\sT
- \frac{J}{J-1} \bar{X}_{\bullet} \bar{X}_{\bullet}^\sT \right] \nonumber \\
&& = \frac{J}{J-1}E[\vec{X}_j\vec{X}_j^\sT]
- \frac{1}{J(J-1)}\left(  \sum_{j=1}^J E[\vec{X}_j\vec{X}_j^\sT ]
+ \sum_{j\ne j^{\prime}}^J E[ \vec{X}_j \vec{X}_{j^\prime}^\sT] \right) \nonumber \\
&& = \frac{J}{J-1}(\bSig + \vec{\mu}\vec{\mu}^\sT ) - \frac{1}{J-1} (\bSig + \vec{\mu}\vec{\mu}^\sT )
- (\bSig^{\prime} + \vec{\mu}\vec{\mu}^\sT ) \nonumber \\
&& = \bSig-\bSig^{\prime} . \hspace*{91mm} \qed \nonumber
\end{eqnarray}

%--------------------------------------------------------------------------------------------------------------- Lemma 2
\vspace*{2mm}
{\it Lemma 2}: Let $\vec{Z} \sim (\vec{0},\bI )$ be independent of the vector-matrix pair $(\vec{A},\bB)$.
Then $V[\vec{A}+\bB \vec{Z}] = V[\vec{A}] + E[\bB \bB^\sT]$.

\vspace*{2mm}
{\it Proof}: 
\begin{eqnarray}
\hspace*{16mm} V[\vec{A}+\bB \vec{Z}] \hspace*{-6mm}
&& = V[E[ \vec{A}+\bB \vec{Z} |\bB]]+E[V[ \vec{A}+\bB \vec{Z} |\bB ]]  \nonumber \\
&& = V[E[ \vec{A}|\bB]+\bB E[\vec{Z}]]+E[V[ \vec{A}|\bB]+\bB V[\vec{Z}] \bB^\sT ]  \nonumber \\
&& = V[E[ \vec{A}|\bB]+\bB \vec{0}]+E[V[ \vec{A}|\bB]+\bB \bI \bB^\sT ]  \nonumber \\
&& = V[E[ \vec{A}|\bB]]+E[V[ \vec{A}|\bB]]+E[\bB \bB^\sT ]  \nonumber \\
&& = V[ \vec{A}]+E[\bB \bB^\sT ]  . \hspace*{55mm} \qed \nonumber
\end{eqnarray}

%--------------------------------------------------------------------------------------------------------------- Lemma 3
\vspace*{2mm}
{\it Lemma 3}: Suppose $\vec{Z}_1$, $\vec{Z}_2$, and $(\vec{A}_1,\vec{A}_2,\bB )$ are mutually
independent with $\vec{Z}_1,\vec{Z}_2\sim (\vec{0},\bI )$. Then
$Cov[\vec{A}_1+\bB\vec{Z}_1,\vec{A}_2+\bB\vec{Z}_2]=Cov[\vec{A}_1,\vec{A}_2]$. 

\vspace*{2mm}
{\it Proof}: 
\begin{eqnarray}
\hspace*{22mm} Cov[\vec{A}_1+\bB \vec{Z}_1,\vec{A}_2+\bB \vec{Z}_2] \hspace*{-30mm} && \nonumber \\
&& = Cov[E[\vec{A}_1+\bB \vec{Z}_1|\bB ],E[\vec{A}_2+\bB \vec{Z}_2|\bB ]]   \nonumber \\
&& \hspace*{10mm} + \, E[ Cov[\vec{A}_1+\bB \vec{Z}_1,\vec{A}_2+\bB \vec{Z}_2|\bB ]] \nonumber \\
&& = Cov[E[\vec{A}_1|\bB ],E[\vec{A}_2|\bB ]]  \nonumber \\
&& \hspace*{10mm} + \, E[ Cov[\vec{A}_1,\vec{A}_2|\bB ]] \nonumber \\
&& = Cov[\vec{A}_1,\vec{A}_2]  . \hspace*{55mm} \qed \nonumber
\end{eqnarray}

%--------------------------------------------------------------------------------------------------------------- Lemma 4
\vspace*{2mm}
{\it Lemma 4}: Let $\vec{S},\vec{S}^\prime$ be independent, identically distributed random vectors independent of the
random vector $\vec{Y}$. Let $F(\vec{Y},\vec{S})$ be a vector function of $\vec{Y}$ and $\vec{S}$. Then
$Cov[F(\vec{Y},\vec{S}),F(\vec{Y},\vec{S}^\prime)] = V[E[F(\vec{Y},\vec{S})|\vec{Y}]]$.

\vspace*{2mm}
{\it Proof}: 
\begin{eqnarray}
Cov[F(\vec{Y},\vec{S}),F(\vec{Y},\vec{S}^\prime)] \hspace*{-30mm} && \nonumber \\
&& = Cov[E[F(\vec{Y},\vec{S})|\vec{Y}],E[F(\vec{Y},\vec{S}^\prime)|\vec{Y}]]   \label{eq:lem4} \\
&& \hspace*{10mm} + \, E[Cov[F(\vec{Y},\vec{S}),F(\vec{Y},\vec{S}^\prime)|\vec{Y}]] . \nonumber 
\end{eqnarray}
\noindent
The two functions $E[F(\vec{Y},\vec{S})|\vec{Y}=\vec{y}]$ and $E[F(\vec{Y},\vec{S}^\prime)|\vec{Y}=\vec{y}]$
are identical so the first covariance on the right in (\ref{eq:lem4}) is $V[E[F(\vec{Y},\vec{S})|\vec{Y}]]$. Also,
$F(\vec{Y},\vec{S})$ and $F(\vec{Y},\vec{S}^\prime)$ are conditionally independent given $\vec{Y}$, so their
conditional covariance on the right in (\ref{eq:lem4}) is zero. \hfill $\qed$

\end{document}